\documentclass[fleqn,usenatbib]{mnras}

\usepackage{newtxtext,newtxmath}
\usepackage[T1]{fontenc}
\usepackage{ae,aecompl}
\usepackage{placeins}
\usepackage{graphicx}	 
\usepackage{amsmath}	
\usepackage{color}
\usepackage{gensymb}

\usepackage{scalerel,tikz}
\usetikzlibrary{svg.path}
\definecolor{orcidlogocol}{HTML}{A6CE39}
\tikzset{orcidlogo/.pic={
 \fill[orcidlogocol] svg{M256,128c0,70.7-57.3,128-128,128C57.3,256,0,198.7,0,128C0,57.3,57.3,0,128,0C198.7,0,256,57.3,256,128z};
 \fill[white] svg{M86.3,186.2H70.9V79.1h15.4v48.4V186.2z}
 svg{M108.9,79.1h41.6c39.6,0,57,28.3,57,53.6c0,27.5-21.5,53.6-56.8,53.6h-41.8V79.1z M124.3,172.4h24.5c34.9,0,42.9-26.5,42.9-39.7c0-21.5-13.7-39.7-43.7-39.7h-23.7V172.4z}
 svg{M88.7,56.8c0,5.5-4.5,10.1-10.1,10.1c-5.6,0-10.1-4.6-10.1-10.1c0-5.6,4.5-10.1,10.1-10.1C84.2,46.7,88.7,51.3,88.7,56.8z};
}}
\newcommand\orcidicon[1]{\href{https://orcid.org/#1}{\mbox{\scalerel*{
\begin{tikzpicture}[yscale=-1,transform shape]
\pic{orcidlogo};
\end{tikzpicture}
}{|}}}}

\newcommand{\aref}[1]{\hyperref[#1]{Appendix~\ref{#1}}}
\defcitealias{2019ApJ...886...14F}{F19}
\defcitealias{2021MNRAS.500.1721M}{M21}

\usepackage[normalem]{ulem}

\definecolor{darkgreen}{rgb}{0.13, 0.55, 0.13}
\definecolor{brown}{rgb}{0.65, 0.16, 0.16}

\DeclareMathOperator\M{\mathcal{M}}

\newcommand{\diff}[1]{\ensuremath{\operatorname{d}\!{#1}}}
\newcommand\hi{H\,\protect\scaleto{$I$}{1.2ex}\,\,}

\title[Mode of turbulence in LMC N159E]{First extragalactic measurement of the turbulence driving parameter: ALMA observations of the star-forming region N159E in the Large Magellanic Cloud}

\author[P. Sharda et al.]{Piyush Sharda$^{\orcidicon{0000-0003-3347-7094}\,1,2}$\thanks{piyush.sharda@anu.edu.au (PS)},
Shyam H. Menon$^{\orcidicon{0000-0001-5944-291X}\,1}$,
Christoph Federrath$^{\orcidicon{0000-0002-0706-2306}\,1,2}$\thanks{christoph.federrath@anu.edu.au (CF)},
Mark R. Krumholz$^{\orcidicon{0000-0003-3893-854X}\,1,2}$\thanks{mark.krumholz@anu.edu.au (MRK)},
\newauthor James R. Beattie$^{\orcidicon{0000-0001-9199-7771}\,1}$,
Katherine E. Jameson$^{\orcidicon{0000-0001-7105-0994}\,3}$\thanks{Bolton Fellow},
Kazuki Tokuda$^{\orcidicon{0000-0002-2062-1600}\,4,5}$,
Blakesley Burkhart$^{\orcidicon{0000-0001-5817-5944}\,6,7}$,
\newauthor Roland M. Crocker$^{\orcidicon{0000-0002-2036-2426}\,1}$,
Charles J. Law$^{\orcidicon{0000-0003-1413-1776}\,8}$,
Amit Seta$^{\orcidicon{0000-0001-9708-0286}\,1}$,
Terrance J. Gaetz$^{\orcidicon{0000-0002-5115-1533}\,8}$,
Nickolas M. Pingel$^{\orcidicon{0000-0001-9504-7386}\,1}$,
\newauthor Ivo R. Seitenzahl$^{\orcidicon{0000-0002-5044-2988}\,9}$,
Hidetoshi Sano$^{\orcidicon{0000-0003-2062-5692}\,5}$, and
Yasuo Fukui$^{\orcidicon{0000-0002-8966-9856}\,10,11}$\\
$^{1}$Research School of Astronomy and Astrophysics, Australian National University, Canberra, ACT 2611, Australia\\
$^{2}$Australian Research Council Centre of Excellence for All Sky Astrophysics in 3 Dimensions (ASTRO 3D), Australia\\
$^{3}$CSIRO, Space and Astronomy, Kensington, WA 6151, Australia\\
$^{4}$Department of Physical Science, Graduate School of Science, Osaka Prefecture University, Sakai, Osaka 599-8531, Japan\\
$^{5}$National Astronomical Observatory of Japan, National Institutes of Natural Science, Mitaka, Tokyo 181-8588, Japan\\
$^{6}$Department of Physics and Astronomy, Rutgers University, Piscataway, NJ 08854, USA\\
$^{7}$Center for Computational Astrophysics, Flatiron Institute, New York, NY 10010, USA\\
$^{8}$Center for Astrophysics \textbar\, Harvard \& Smithsonian, Cambridge, MA 02138, USA\\
$^{9}$School of Science, University of New South Wales, Australian Defence Force Academy, Canberra, ACT 2600, Australia\\
$^{10}$Department of Physics, Nagoya University, Chikusa-ku, Nagoya 464-8602, Japan\\
$^{11}$Institute for Advanced Research, Nagoya University, Chikusa-ku, Nagoya 464-8601, Japan}

\date{Accepted 2021 October 18. Received 2021 October 18; in original form 2021 September 8}

\pubyear{2021}

\begin{document}
\label{firstpage}
\pagerange{\pageref{firstpage}--\pageref{lastpage}}
\maketitle

\begin{abstract}
Studying the driving modes of turbulence is important for characterizing the impact of turbulence in various astrophysical environments. The driving mode of turbulence is parameterized by $b$, which relates the width of the gas density PDF to the turbulent Mach number; $b\approx 1/3$, $1$, and $0.4$ correspond to driving that is solenoidal, compressive, and a natural mixture of the two, respectively. In this work, we use high-resolution (sub-pc) ALMA $^{12}$CO ($J$~=~2--1), $^{13}$CO ($J$~=~2--1), and C$^{18}$O ($J$~=~2--1) observations of filamentary molecular clouds in the star-forming region N159E (the Papillon Nebula) in the Large Magellanic Cloud (LMC) to provide the first measurement of turbulence driving parameter in an extragalactic region. We use a non-local thermodynamic equilibrium (NLTE) analysis of the CO isotopologues to construct a gas density PDF, which we find to be largely log-normal in shape with some intermittent features indicating deviations from lognormality. We find that the width of the log-normal part of the density PDF is comparable to the supersonic turbulent Mach number, resulting in $b \approx 0.9$. This implies that the driving mode of turbulence in N159E is primarily compressive. We speculate that the compressive turbulence could have been powered by gravo-turbulent fragmentation of the molecular gas, or due to compression powered by \hi flows that led to the development of the molecular filaments observed by ALMA in the region. Our analysis can be easily applied to study the nature of turbulence driving in resolved star-forming regions in the local as well as the high-redshift Universe. 
\end{abstract}


%%%%%%%%%%%%%%%%% BODY OF PAPER %%%%%%%%%%%%%%%%%%

\section{Introduction}
\label{s:intro}
Turbulence is ubiquitous in the Universe and plays an important role in most astrophysical environments, from stellar interiors \citep{2009AnRFM..41..317M} to galaxy clusters \citep{2014Natur.515...85Z}. In many such environments, the impact of turbulence on the evolution of the gas is very complex \citep[][]{2020ApJ...905...14B}. In particular, several works have shown how turbulence non-linearly impacts the physics of star formation \citep[see recent reviews by][]{2020SSRv..216...62R,2020SSRv..216...68G}. A classic example is that turbulence can either assist or hinder star formation in the interstellar medium (ISM) that can lead to an order of magnitude difference in star formation rates depending on the turbulence driving source \citep{2012ApJ...761..156F,2015ApJ...806L..36S}. This is because the driving mode of turbulence influences the dynamics of the star-forming gas: if the turbulence is primarily composed of compressive (curl-free) modes, it will aid the compression of the gas, thus leading to the formation of dense cores. On the other hand, if the turbulence is primarily composed of solenoidal (divergence-free) modes, gas compression is significantly reduced compared to compressive driving \citep{2008ApJ...688L..79F,2010A&A...512A..81F}. The driving mode of turbulence sets the density and velocity fluctuations in star-forming molecular clouds, which in turn control collapse and fragmentation \citep{2005ApJ...630..250K,2012ApJ...761..156F,2013ApJ...763...51F,2013A&A...553L...8K,2015ApJ...808...48B,2017A&A...608L...3K}. It also influences the chemistry and thermodynamics as the clouds collapse \citep{1999ApJ...527..673S,2009ApJ...692..594P,2016A&A...595A..94I,2019MNRAS.490..513S,2020MNRAS.497..336S,2021MNRAS.503.2014S,2020MNRAS.493.3098M}. Thus, characterizing the mode of turbulence is critical to further our understanding of the lifecycle of molecular clouds.

There has been immense progress towards characterizing the driving mode of turbulence in the ISM, both in theory and simulations \citep[e.g.,][]{2010A&A...512A..81F,2016ApJ...825...30P,2017MNRAS.469..383J,2017MNRAS.472.2496K,2020MNRAS.493.3098M,2020MNRAS.493.4643M,2020ApJ...893...75L}. In the absence of magnetic fields and assuming an isothermal equation of state, the mode of turbulence is typically defined as the ratio of the width of the normalized gas density probability distribution function (assumed to be log-normal), $\sigma_{\rho/\rho_0}$, to the turbulent Mach number, $\mathcal{M}$ \citep{1997MNRAS.288..145P,1998PhRvE..58.4501P}\footnote{Note that this relation is valid in the case of supersonic turbulence ($\mathcal{M} > 1$). In subsonic turbulence there are no compressive shocks and disturbances are primarily propagated by sound waves \citep{2019MNRAS.484.4881M,2020MNRAS.493.5838M}. However, \autoref{eq:b} is often still a good approximation even in the subsonic case \citep{2012ApJ...761..149K,2015MNRAS.451.1380N}.}
\begin{equation}
    b = \frac{\sigma_{\rho/\rho_0}}{\mathcal{M}}\,.
\label{eq:b}
\end{equation}
Simulations find that fully solenoidal driving of the turbulence produces $b\approx 1/3$, fully compressive driving produces $b\approx 1$, and a natural (steady-state) mixture of the two yields $b \approx 0.4$ \citep{2010A&A...512A..81F,2011PhRvL.107k4504F}. The physical origin of this relationship is simple: a driving field that contains primarily compressive modes produces stronger compressions and rarefactions, and thus results in a higher spread in the density probability distribution function (PDF) than a primarily solenoidal velocity field \citep[see, for instance,][]{2008ApJ...688L..79F,2012ApJ...755L..19B,2021MNRAS.504.4354B}. Examples of driving mechanisms that are largely compressive include hydrodynamical shocks, spiral waves, ionizing feedback from massive stars \citep{2017IAUS..322..123F,2020MNRAS.493.4643M,2021MNRAS.504.4354B}. On the other hand, solenoidal turbulence is believed to be driven by magnetorotational instabilities (MRI) in accretion discs, protostellar outflows and stellar winds, and shearing motions, such as those induced by differential rotation \citep{2012ApJ...747...22H,2014ApJ...784...61O,2016ApJ...832..143F,2017ApJ...847..104O,2018NatAs...2..896O}. It is also important to note that if strong magnetic fields are present, then $\mathcal{M} \to \mathcal{M}(1+1/\beta)^{-1/2}$ in \autoref{eq:b}. Here, $\beta$ is the turbulent plasma beta \citep[e.g.,][]{2012ApJ...761..156F} that describes the strength of thermal to magnetic pressure: $\beta = 2\mathcal{M}^2_{\rm{A}}/\mathcal{M}^2$, where $\mathcal{M}_{\rm{A}}$ is the Alfvén Mach number. The driving mode can also vary from between solenoidal and compressive both in time and space as the molecular clouds evolve \citep{2017MNRAS.472.2496K,2017A&A...599A..99O,2020MNRAS.497.1263K}. Thus, $b$ is best conceived as representing the \textit{instantaneous} driving mode of turbulence in a system. 

Despite great progress in measuring the driving mode of turbulence in simulations, it has only been investigated in a very limited sample of observations (\citealt{1997ApJ...474..730P,2010A&A...513A..67B,2013ApJ...779...50G,2016ApJ...832..143F,2017A&A...608L...3K,2020arXiv201203160M,2021MNRAS.500.1721M}; hereafter, \citetalias{2021MNRAS.500.1721M}), all of which remain limited to the Milky Way. This is because measuring $b$ requires constructing the gas density PDF, which needs either dust observations spanning a wide range of frequencies and extinctions \citep[e.g.,][]{2013ApJ...766L..17S,2015MNRAS.453L..41S}, or multiple, optically-thin gas density tracers for gas at different densities \citepalias[e.g.,][]{2021MNRAS.500.1721M}. Moreover, measuring $b$ also requires high spatial resolution from which reliable gas kinematics can be obtained \citep[e.g.,][]{2018MNRAS.477.4380S}. These issues have so far prevented us from measuring the driving mode of turbulence in extragalactic observations.

In this paper, we provide the first observational measurement of the driving mode of turbulence in an extragalactic region. We utilize Atacama Large Millimeter/Submillimeter Array (ALMA) observations of the star-forming region N159E in the Large Magellanic Cloud (LMC) that contains the Papillon Nebula \citep[hereafter, \citetalias{2019ApJ...886...14F}]{2019ApJ...886...14F}. Located southwest of the core of the starburst region 30 Doradus, N159E lies at the eastern end of the star-forming giant molecular cloud (GMC) N159 \citep{1994A&A...291...89J,2008ApJS..178...56F,2008ApJS..175..485M,2010ApJ...721.1206C}. The exquisite resolution provided by ALMA has enabled spatially-resolved molecular gas studies of different parts of N159 on sub-pc scales in great detail \citep[e.g.,][]{2015ApJ...807L...4F,2017ApJ...835..108S,2018ApJ...854..154N,2019ApJ...886...15T}. Since the discovery of the first extragalactic protostar in the region \citep{1981MNRAS.197P..17G}, further observations have revealed that N159 hosts a number of massive young stellar objects \citep[YSOs;][]{1999A&A...352..665H,2004AJ....128.2206I,2004A&A...422..129M,2005AJ....129..776N,2006A&A...453..517T,2007A&A...469..459T,2009ApJ...699..150S,2010ApJ...721.1206C,2017ApJ...835..108S,2018ApJ...854..154N,2020A&A...643A..63G}, making it an ideal region to study the turbulence in molecular clouds impacted by protostellar feedback. In this work, we measure the mode of turbulence in the filamentary structure comprised of molecular gas, including outflows from two newly-identified dense cores that are suspected to harbour massive, young ($< 10^4\,\mathrm{yr}$) protostars in N159E \citepalias{2019ApJ...886...14F}.

We arrange the rest of this paper as follows: \autoref{s:data} describes the ALMA data we use in this work, \autoref{s:analysis} describes how we measure the gas density PDF and the turbulent Mach number, \autoref{s:resultsdiscussions} presents our results, \autoref{s:caveats} lays out the caveats for our work, and \autoref{s:possibledrivers} discusses the possible drivers of turbulence in N159E. Finally, we present our conclusions in \autoref{s:conclusions}. For this work, we assume the metallicity of the LMC to be $0.5\,\rm{Z_{\odot}}$ \citep[e.g.,][]{1998ApJ...505..732H,2016AJ....151..161S}, and distance to the LMC to be $50\,\mathrm{kpc}$ \citep[e.g.,][]{2019Natur.567..200P}, corresponding to 0\farcs1 $= 0.24\,\mathrm{pc}$.

\section{Data}
\label{s:data}
N159E was observed during Cycle 4 using ALMA Band 6 (P.I. Y. Fukui, \#2016.1.01173.S.), targeting multiple molecular lines -- $^{12}$CO $(J=2-1)$, $^{13}$CO $(J=2-1)$, C$^{18}$O $(J=2-1)$, and SiO $(J=5-4)$, as well as the $1.3\,\rm{mm}$ continuum. We refer the readers to \citetalias{2019ApJ...886...14F} and \cite{2019ApJ...886...15T} for details on the data reduction and analysis of these ALMA observations; here, we note the main features that are relevant to this work. The beam size of these observations is 0.28\arcsec $\times$ 0.25\arcsec (for the CO isotopologues\footnote{Isotopologues refer to molecules differing only in the isotope of one or more of the component atoms.}) and 0.26\arcsec $\times$ 0.23\arcsec (for the $1.3\,\rm{mm}$ continuum), corresponding to a spatial resolution of $\sim 0.07\,\rm{pc}$. The observations detected significant flux in all three CO lines, at an rms noise level of $\sigma_{^{12}\rm{CO}} \sim 4\,\rm{mJy\,beam^{-1}}$ for $^{12}$CO, $\sigma_{^{13}\rm{CO}}\sim 4.5\,\rm{mJy\,beam^{-1}}$ for $^{13}$CO, and $\sigma_{\rm{C^{18}O}}\sim 4.5\,\rm{mJy\,beam^{-1}}$ for C$^{18}$O. These observations also revealed detailed, filamentary structures comprised of molecular gas that were unresolved in previous Cycle 1 ALMA observations of N159E at lower resolution ($\sim 0.24\,\rm{pc}$, \citealt{2017ApJ...835..108S}). We do not utilize the Cycle 1 data as the Cycle 4 data shows no significant missing flux at higher resolution \citepalias{2019ApJ...886...14F}.

\autoref{fig:fullmap} shows the integrated intensity (moment 0) map of $^{13}$CO in N159E as imaged by ALMA (see also, figure~1 of \citetalias{2019ApJ...886...14F}). The Papillon Nebula YSO binary system (masses $\sim 21\,\rm{M_{\odot}}$ and $41\,\rm{M_{\odot}}$ -- \citealt{1999A&A...352..665H,2004AJ....128.2206I,2007A&A...469..459T,2010ApJ...721.1206C}) lie towards the southern half of the region, marked by the magenta box. The black box encloses the $\sim 1\times4\,\rm{pc}$ region we analyze in this work (hereafter, referred to as the `analysis region'). It possibly consists of filamentary molecular gas structures (see figure~4 of \citetalias{2019ApJ...886...14F}) that trace a larger-scale \hi gas distribution \citep{2017PASJ...69L...5F}, and two other compact sources called MMS-1 and MMS-2 identified by \citetalias{2019ApJ...886...14F}. These compact sources are most likely massive YSOs believed to have formed as a result of colliding flows (\citetalias{2019ApJ...886...14F}; \citealt{2021PASJ...73S...1F}), which are responsible for driving outflows within the region. The outflowing gas spans $\sim 30\,\rm{km\,s^{-1}}$ in velocity. Note that we cannot put constraints on the actual size of the outflows (although they are believed to be $0.1\,\rm{pc}$ long), or distinguish between the outflowing gas from one massive YSO to the other. Thus, our analysis region consists of a mixture of filamentary molecular clouds where massive YSOs driving outflows are present. We cannot exclude a part of the contiguous structure in this region because a sufficiently large field of view is necessary to properly sample the low column density part of the PDF \citep{2019MNRAS.482.5233K}, as well as to retrieve reliable gas kinematics from the data \citep{2018MNRAS.477.4380S}.

\section{Analysis}
\label{s:analysis}
Below we describe the general method we adopt to derive the column density and velocity structure of molecular gas in the region, construct their PDFs, and calculate their respective density and velocity dispersions.

\subsection{Gas density PDF}
\label{s:densityPDF}
We use the three CO isotopologues to construct the gas density PDF. We also assume the following abundance ratios for N159E: $X[\rm{H_2}/^{12}\mathrm{CO}]=62500$ \citep{2008ApJS..178...56F}, $X[^{12}\mathrm{CO}/^{13}\mathrm{CO}]=50$ (\citealt{2010PASJ...62...51M}; \citetalias{2019ApJ...886...14F}), and $X[^{12}\mathrm{CO}/\mathrm{C^{18}O}]=560$ \citep{2009ApJ...690..580W}, and we discuss the effects of varying these abundance ratios later in \autoref{s:caveats}. Below, we first estimate the column density while assuming the gas to be in local thermodynamic equilibrium (LTE), as is commonly done in such analyses \citepalias[e.g.,][]{2021MNRAS.500.1721M}. Then, we relax this assumption and perform a non-LTE (NLTE) analysis to derive the gas density PDF, showing how sub-thermal excitation of $^{13}$CO, in particular, can give rise to important deviations from an LTE approach for star-forming molecular clouds.

\begin{figure}
\includegraphics[width=1.0\columnwidth]{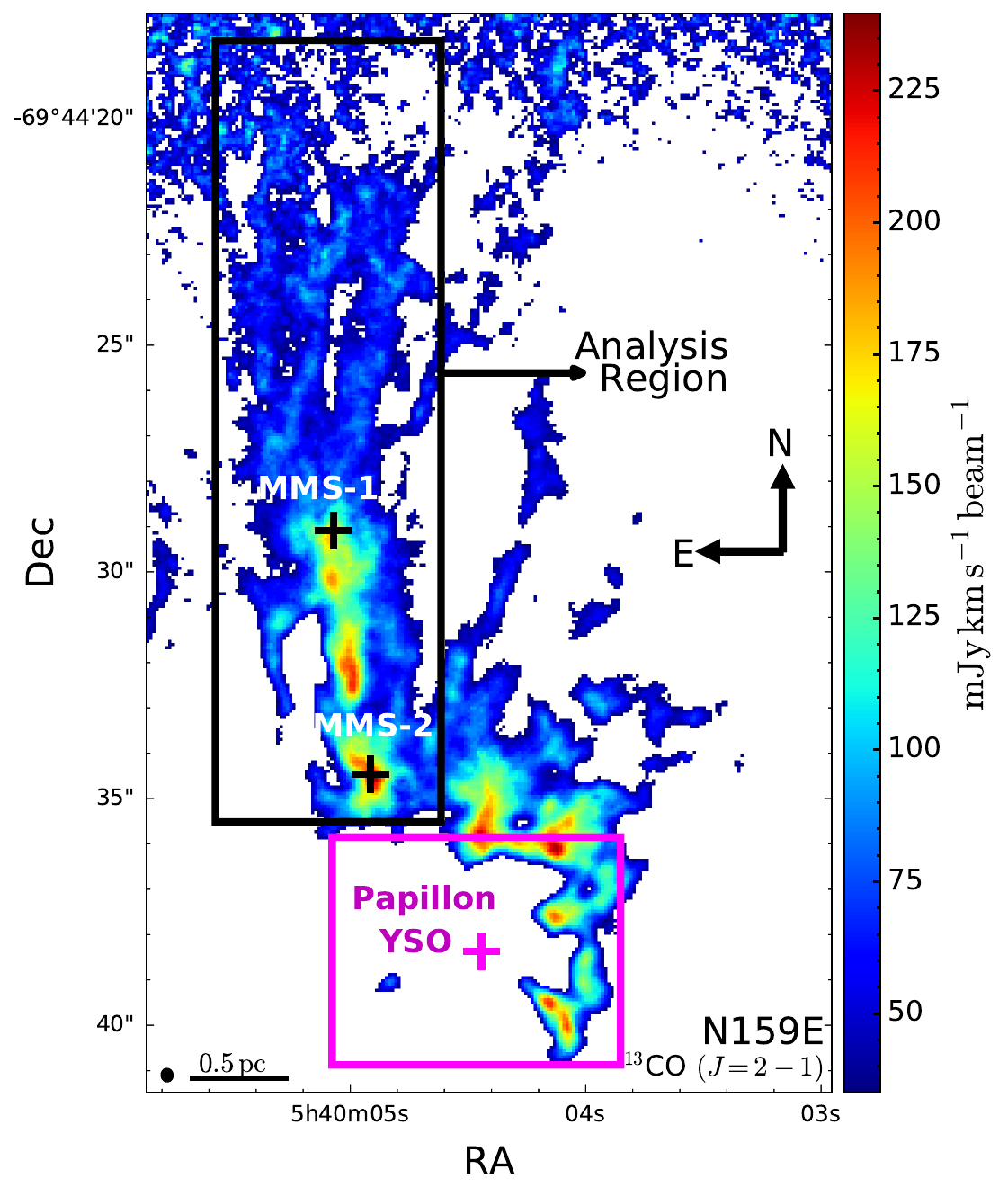}
\caption{ALMA $^{13}$CO ($J$~=~2--1) integrated intensity map of N159E in the LMC. The magenta box marks the approximate extent of the Papillon Nebula that harbours the Papillon YSO binary system marked by the magenta cross. The $1\times4\,\rm{pc}$ analysis region we use to measure the driving mode of turbulence is denoted by the solid, black box. This region contains two young, dense cores marked by the black crosses (MMS-1 and MMS-2, as identified by \protect\citetalias{2019ApJ...886...14F}) that are suspected to contain massive YSOs that are driving $\sim 0.1\,\rm{pc}$ sized outflows in this region. The beam size (angular resolution $\approx 0.28\arcsec \times 0.25\arcsec$) is represented by the black circle in the bottom left corner, and the adjacent scale shows the physical extent of the region. The noisiness in the pixels towards the North is caused by the drop in sensitivity at the edge of the mosaic.}
\label{fig:fullmap}
\end{figure}

\begin{figure*}
\includegraphics[width=1.0\columnwidth]{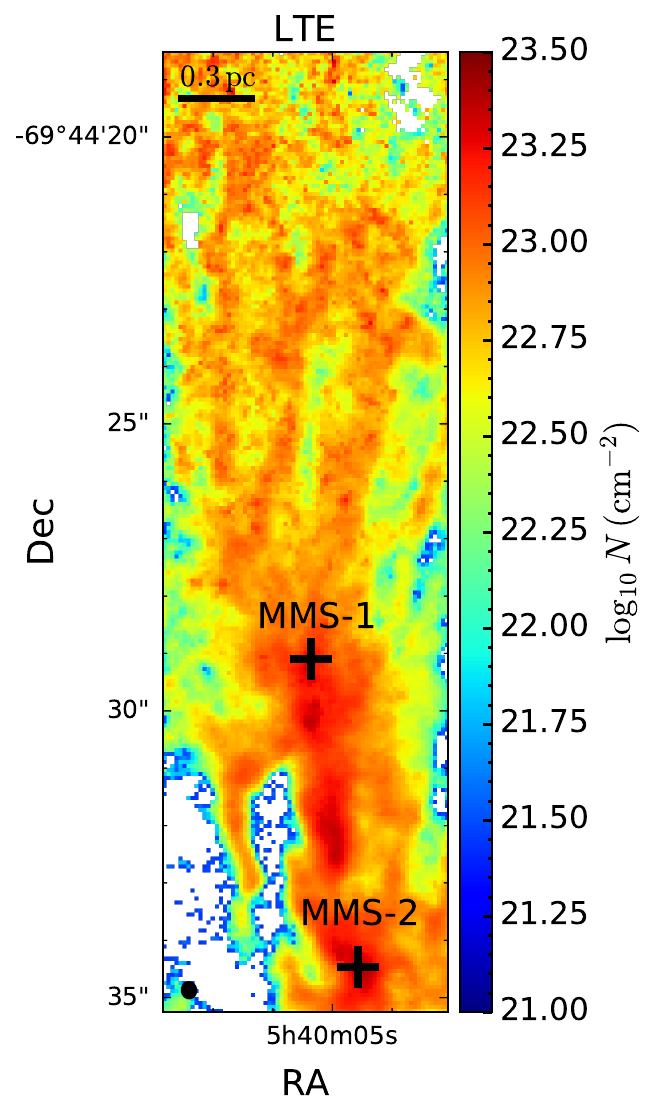}
\includegraphics[width=1.0\columnwidth]{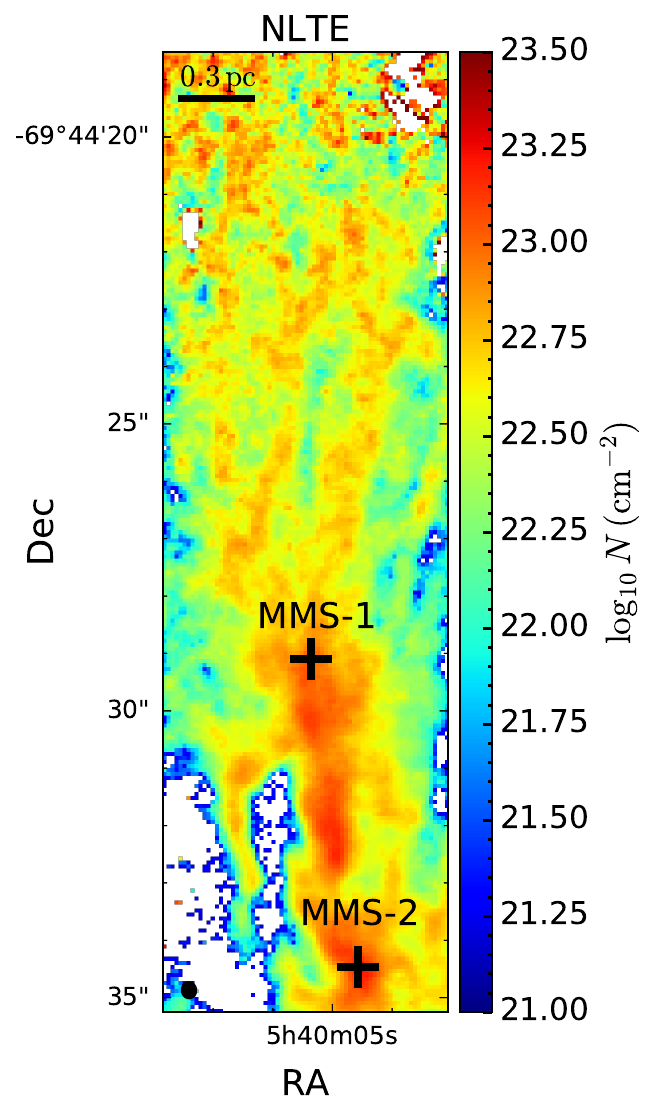}
\caption{H$_2$ column density map of the analysis region shown in \autoref{fig:fullmap}, created using the $^{12}$CO ($J$~=~2--1), $^{13}$CO ($J$~=~2--1), and C$^{18}$O ($J$~=~2--1) ALMA observations. The left panel shows the resulting column density when assuming CO isotopologues to be in local thermodynamic equilibrium (LTE) whereas the right panel shows the map for a non-LTE (NLTE) analysis. The beam size and scale are denoted in the bottom left and the top left corner of the plot, respectively. Black crosses denote the probable positions of the two massive, young protostars (MMS-1 and MMS-2) that are driving outflows in this region \protect\citepalias{2019ApJ...886...14F}.}
\label{fig:columndensitymap}
\end{figure*}

\begin{figure}
\includegraphics[width=1.0\columnwidth]{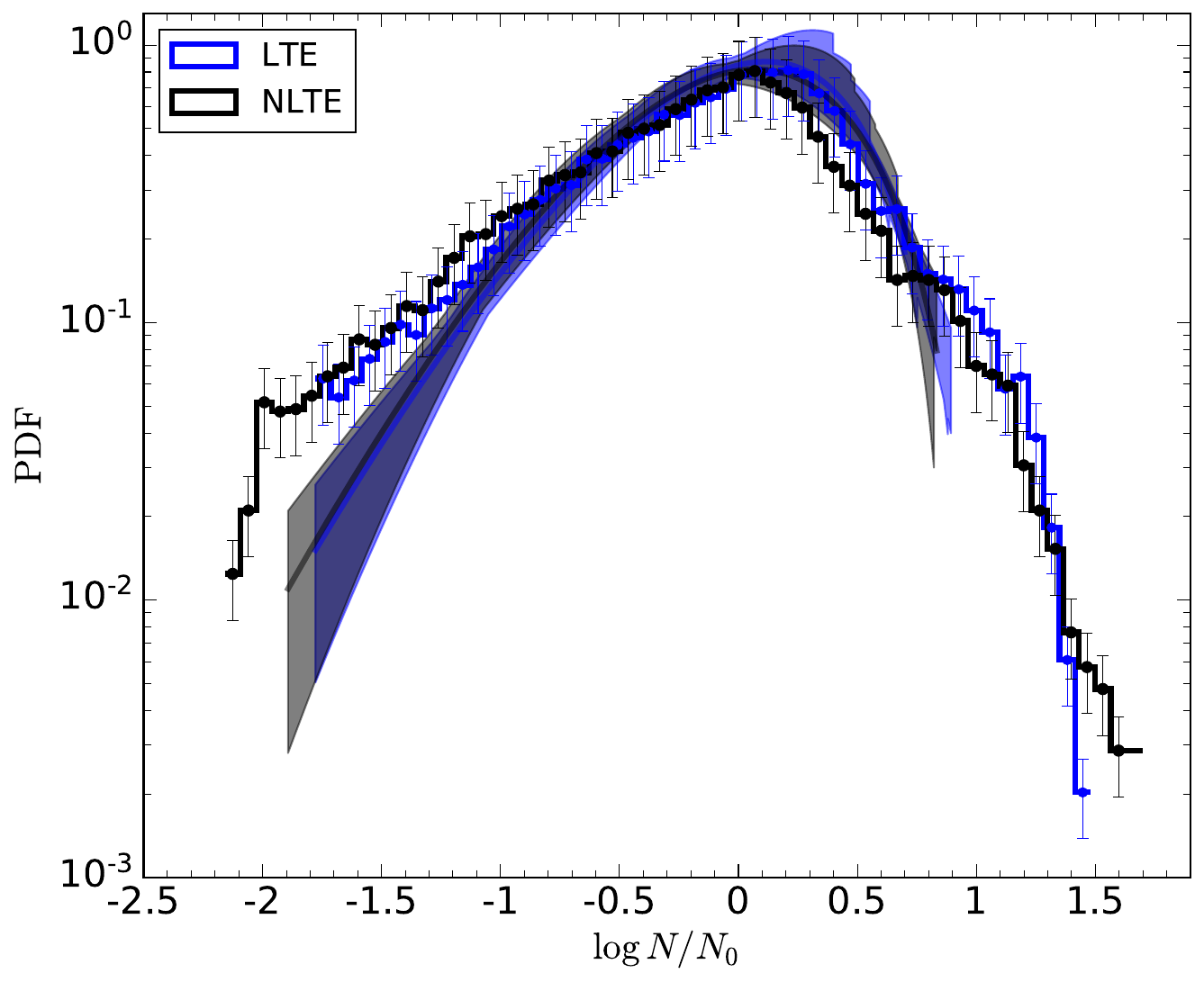}
\caption{Normalized 2D column density PDF (with $1\sigma$ errors) of the analysis region shown in \autoref{fig:columndensitymap}, using the LTE and the NLTE methods described in \autoref{s:densityPDF}. $N_0$ represents the mean column density. The distribution is largely log-normal, with some intermittent features indicating deviations from non lognormality. Fitting the \protect\citet{Hopkins_2013} gas density PDF model (solid lines represent the best fit and shaded curves represent the $1\sigma$ error on the best fit) gives the 2D column density dispersion in the LTE case $\sigma_{N/N_0}\rm{(LTE)}=0.58\pm0.14$, and in the NLTE case $\sigma_{N/N_0}\rm{(NLTE)}=0.51\pm0.04$. The corresponding intermittencies are $\theta = 0.12\pm0.05$ (LTE) and $\theta = 0.17\pm0.09$ (NLTE). We use these 2D column density dispersions to derive the 3D density dispersion $\sigma_{\rho/\rho_0}$, following \protect\cite{Brunt_2010b}.}
\label{fig:coldumndensitypdf}
\end{figure}

\begin{figure*}
\includegraphics[width=1.0\linewidth]{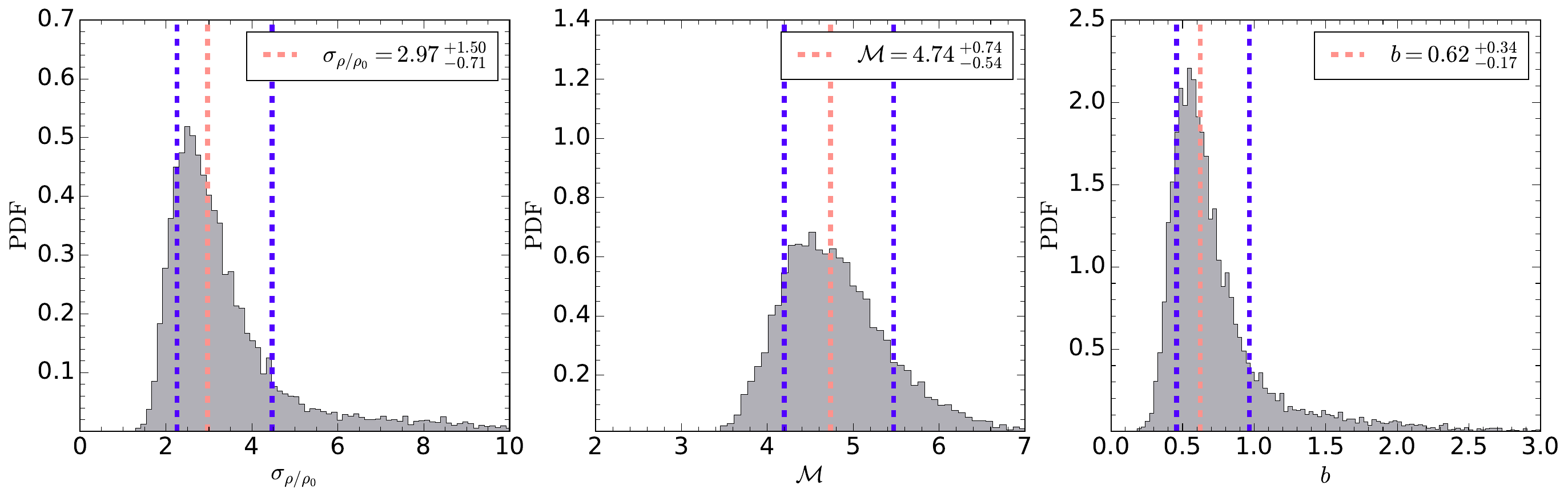}
\includegraphics[width=1.0\linewidth]{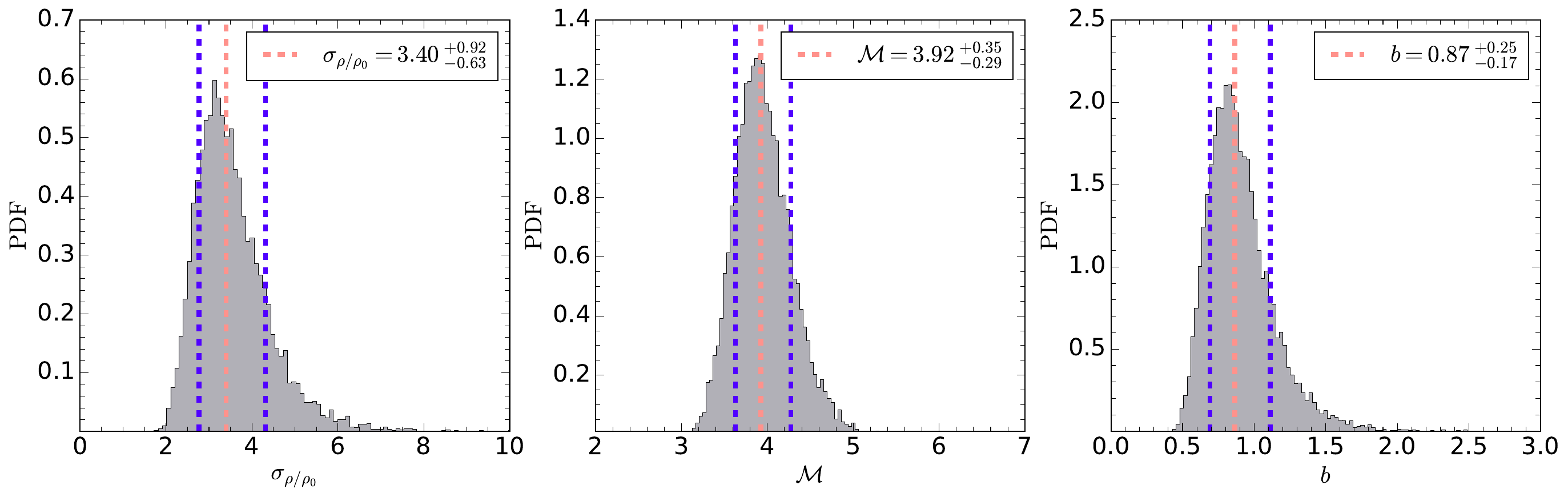}
\caption{PDFs of the distributions of the 3D density dispersion ($\sigma_{\rho/\rho_0}$), turbulent Mach number ($\mathcal{M}$), and the driving mode of turbulence ($b$) obtained from 10,000 bootstrapping realisations. The top and the bottom row corresponds to the LTE and the NLTE analyses, respectively. The dashed orange line denotes the $50^{\rm{th}}$ percentile, whereas the dashed blue lines denote the $16^{\rm{th}}$ and the $84^{\rm{th}}$ percentiles, respectively.}
\label{fig:distributions}
\end{figure*}

\subsubsection{LTE analysis}
\label{s:lte}
To trace the $\mathrm{H}_2$ gas column density structure under the LTE approximation, we follow \citetalias{2021MNRAS.500.1721M} and use a hybrid method with a combination of the $^{13}$CO and C$^{18}$O ($J$~=~2--1) lines. We do not use the $^{12}\mathrm{CO}$ emission to directly trace the $\mathrm{H}_2$ column density in this work, primarily because we expect $^{12}\mathrm{CO}$ to be optically thick in most of the analysis region, so it cannot accurately trace the true column density of $\mathrm{H}_2$ \citep[however, see][for an alternative method with a correction factor for optically-thick lines]{2015PASP..127..266M}. In the hybrid method, we use C$^{18}$O emission to trace the densest parts of the region where $^{13}$CO becomes optically thick, otherwise we use the $^{13}$CO emission. We estimate the optical depth of $^{13}$CO as $\tau_{13} = \tau_{12}X[^{13}\mathrm{CO}/^{12}\mathrm{CO}]$, where $\tau_{12}$ is the optical depth of $^{12}\mathrm{CO}$ that we calculate using the line ratio $^{\rm{iso}}R_2$ ($^{12}$CO ($J$~=~2--1) to $^{13}$CO ($J$~=~2--1)) following equation 1 of \citet{Choi_1993}. Thus, we can determine from this relation whether $^{13}\mathrm{CO}$ is optically thick (\textit{i.e.,} $\tau_{13}>1$) in a given spectral channel and spatial region, in which case we use $\mathrm{C}^{18}\mathrm{O}$ to trace the $\mathrm{H}_2$ column density. We find that the maximum optical depths in the analysis region are such that C$^{18}$O is never measured to become optically thick.

We then obtain the H$_2$ column density contribution for each pixel (with more than $3\sigma_{\rm{^{13}CO}}$ emission) from the main-beam brightness temperature $T_\mathrm{MB}$ of the relevant line as
\begin{equation}
\label{eq:Column_Density}
N_{\mathrm{H}_2} = X [\mathrm{H}_2] \int f(T_\mathrm{ex})  T_\mathrm{MB} \, dv ,
\end{equation}
where $ X [\mathrm{H}_2]$ is the abundance factor used to convert the line column density to a H$_2$ column density, and $f(T_\mathrm{ex})$ is a function of the excitation temperature ($T_{\mathrm{ex}}$). The abundance factor is given by $ X [\mathrm{H}_2] = X [^{12}\mathrm{CO}/^{13}\mathrm{CO}] \, X[\mathrm{H}_2/^{12}\mathrm{CO}]$ and $X [^{12}\mathrm{CO}/\mathrm{C} ^{18} \mathrm{O}] \, X[\mathrm{H}_2/^{12}\mathrm{CO}]$ for $^{13}\mathrm{CO}$ and $\mathrm{C}^{18}\mathrm{O}$, respectively. The function $f$ is given by \citep{2015PASP..127..266M}
\begin{equation}
\label{eq:Prefactor}
	f \left(T_{\mathrm{ex}} \right) = \frac{3hZ}{8 \pi^3 \mu^2 J_u} \frac{\exp\left(\frac{E_\mathrm{up}}{kT_{\mathrm{ex}}}\right) }{1-\exp \left( \frac{h\nu}{kT_{\mathrm{ex}}} \right)} \frac{1}{J \left(T_{\mathrm{ex}} \right)-J\left(T_\mathrm{BG} \right)} ,
\end{equation}
where $\nu$ is the rest-frame line frequency, $\mu$ the molecule dipole moment, $Z$ the rotational partition function that can be approximated as $Z \approx kT_{\mathrm{ex}}/hB + 1/3$ where $B = \nu/2J_u$ is the rotational constant, $E_\mathrm{up}$ the energy of the upper rotational level, $J_u = 2$ the upper quantum level number, and $J(T_{\mathrm{ex}})-J(T_\mathrm{BG})$ the correction for background CMB radiation, where
\begin{equation}
\label{eq:J_equation}
	J(T) = \frac{h\nu}{k \left(\exp\left(\frac{h\nu}{kT}\right)-1 \right)} .
\end{equation}
Based on the peak brightness temperature of $^{12}$CO $\sim 40\,\rm{K}$ in the region of interest as derived by \citetalias{2019ApJ...886...14F}, we use $T_{\mathrm{ex}} = T_{\mathrm{gas}} = (40\pm10)\,\rm{K}$, under the assumption of LTE. An alternative approach would be to use the $^{12}$CO main-beam brightness temperature ($T_{\mathrm{MB}}$) as an estimate of the excitation temperature, assuming $^{12}$CO is optically thick. However, \citetalias{2021MNRAS.500.1721M} find that this makes a negligible difference to the width of the normalized column density PDF (although the absolute values can change) -- which is the only relevant quantity in the context of our analysis -- as long as the inferred excitation temperatures are in the range $T_\mathrm{ex} \sim (10-70) \, \mathrm{K}$. 

The left panel of \autoref{fig:columndensitymap} shows the resulting H$_2$ column density map that we obtain for the analysis region under the LTE approximation. The mean column density in the region over all pixels with detections is $N_{0,\rm{LTE}} = (6.0 \pm 0.5)\times 10^{22}\,\rm{cm^{-2}}$, higher by a factor of 10 than the median H$_2$ column density in young LMC GMCs \citep{2010MNRAS.406.2065H} but comparable to the H$_2$ column density in the LMC starburst region 30Doradus \citep{2013ApJ...774...73I}. The maximum column density reaches as high as $\sim 3\times10^{23}\,\rm{cm^{-2}}$, indicating the presence of dense cores in the region. We describe the analysis to extract the width of the column density PDF later in \autoref{s:hopkinsPDF}.

\subsubsection{Non-LTE analysis}
\label{s:nonlte}
So far, we have assumed that the excitation temperature of the CO isotopologues is the same as the kinetic temperature of the gas. However, in the case of N159E, given the expected densities \citepalias{2019ApJ...886...14F} and gas kinetic temperatures, we expect $^{13}$CO to be sub-thermally excited \citep[see, for example, figure~10 of][]{1999ApJ...517..209G}. Thus, the assumption of LTE will most likely break down. Additionally, another advantage of following a non-LTE approach is that we do not have to \textit{a priori} assume an excitation temperature for the isotopologues because it is self-consistently determined from a set of best-fit NLTE models that match the data.

We use the large velocity gradient (LVG) approximation for the purpose of estimating column densities without assuming LTE \citep{1960mes..book.....S}. Briefly, the LVG approximation assumes that photon absorptions are local to the sites at which photons are emitted, so the probability that a photon escapes depends on the local velocity gradient at the point of emission \citep[e.g.,][]{1975ApJ...199...69D,1980A&A....91...68D}. We first create a 3D grid of non-LTE models based on the number density of H nuclei ($n_{\rm{H}}$), gas kinetic temperature ($T_{\rm{g}}$) and velocity gradient ($d \mathrm{v}/dr$) along the line of sight using the software library \texttt{DESPOTIC} \citep{2014MNRAS.437.1662K}. The range in $n_{\rm{H}}$ and $T_{\rm{g}}$ covered by the grid is $10^2-10^{6.5}\,\rm{cm^{-3}}$ and $5-100\,\rm{K}$, respectively. Based on the results of \citetalias{2019ApJ...886...14F}, we use the plane-of-sky velocity of $\sim4\,\rm{km\,s^{-1}}$ as an estimate of $d\rm{v}$ along the line of sight, and vary the line of sight depth between $0.01-100\,\rm{pc}$ to construct the grid in $d\mathrm{v}/dr$. The resolution of the grid in $n_{\rm{H}},\,T_{\rm{g}},\,d\mathrm{v}/dr$ is $350\times20\times20$, respectively. We checked for the convergence of the model grid, finding that the results do not change if the grid resolution is further increased in any dimension. We also ensure that the models that best describe the data do not lie along any of the edges in the parameter space; thus, the coverage in density, temperature, and velocity gradient in the model grid is sufficient to derive meaningful physical properties from the data.

For a given set of {$n_{\rm{H}},\,T_{\rm{g}},\,d\mathrm{v}/dr$} and abundances of emitting species (in this case, $^{12}$CO, $^{13}$CO and C$^{18}$O), \texttt{DESPOTIC} gives the excitation temperatures ($T_{\rm{ex}}$), optical depths ($\tau$), and velocity-integrated brightness temperatures ($W$) for every rotational line produced by these species. We assign a $\chi^2$ to every model for a given pixel in the data based on the ratios of the integrated brightness temperatures of $^{12}$CO/$^{13}$CO, and $^{12}$CO/C$^{18}$O
\begin{eqnarray}
    \lefteqn{\chi^2_{\rm{mod,pix}} = \frac{1}{2} \left(\frac{W_{\rm{mod,^{12}CO}}/W_{\rm{mod,^{13}CO}} - W_{\rm{pix,^{12}CO}}/W_{\rm{pix,^{13}CO}}}{\sigma_{\left(W_{\rm{pix,^{12}CO}}/W_{\rm{pix,^{13}CO}}\right)}}\right)^2 +
    }
    \nonumber \\
    & & {}  + \frac{1}{2} \left(\frac{W_{\rm{mod,^{12}CO}}/W_{\rm{mod,C^{18}O}} - W_{\rm{pix,^{12}CO}}/W_{\rm{pix,C^{18}O}}}{\sigma_{\left(W_{\rm{pix,^{12}CO}}/W_{\rm{pix,C^{18}O}}\right)}}\right)^2,\,
    \label{eq:chi2}
\end{eqnarray}
where the subscripts $\rm{mod}$ and $\rm{pix}$ refer to the models and the data in a pixel, respectively. We have tried different combinations of the model properties that can be matched against the data, and find that the ratio of integrated brightness temperatures gives the most reasonable and well-constrained match against the data. This is because using a combination of optically thin and optically thick tracers allows us to overcome the challenges one faces when using only one of the two category of tracers \citep{2013ApJ...771..122B}. Finally, to get the set of column density, gas temperature, velocity gradient, excitation temperatures, and optical depths that best describe the data in a given pixel, we take a $\chi^2$-weighted mean of all the models
\begin{equation}
\begin{split}
     \{N,\,n_{\mathrm{H}},\,T_{\mathrm{g}},\,d \mathrm{v}/dr,\,T_{\rm{ex}},\,\tau\}_{\rm{pix}} = \\
     \frac{\sum_{i=1}^{N_{\mathrm{mod}}} \{N,\,n_{\mathrm{H}},\,T_{\mathrm{g}},\,d \mathrm{v}/dr,\,T_{\mathrm{ex}},\,\tau\}_{i}e^{-\chi^2_{i}}}{\sum_{i=1}^{N_{\rm{mod}}} e^{-\chi^2_{i}}} 
\end{split}\,,
\end{equation}
where $N_{\rm{mod}}=140,000$ is the total number of models in our grid.

The right panel of \autoref{fig:columndensitymap} shows the resulting column density map that we recover from the NLTE analysis. The mean column density in this case is $N_{0,\rm{NLTE}} = (4.17\pm0.42)\times 10^{22}\,\rm{cm^{-2}}$, slightly lower than what we find from the LTE analysis above. The best-fit means (over all pixels with detections) are $n_{\rm{H}} = (1.72\pm1.24)\times10^{5}\,\rm{cm^{-3}}$, $T_{\rm{g}}=58\pm8\,\rm{K}$ and $d \mathrm{v}/dr = (1.33\pm0.72)\times10^{-12}\,\rm{s^{-1}}$. The mean gas volume density is a factor $2-6$ smaller than that found in the dense pillars around the Papillon Nebula that lie south of the analysis region \citepalias[][table~1]{2019ApJ...886...14F}. Both the best-fit mean $n_{\rm{H}}$ and $T_{\rm{g}}$ are in good agreement with measurements of the density and the gas kinetic temperature in other massive star-forming regions in the LMC \citep[e.g.,][]{2017A&A...600A..16T,2021arXiv210810519T}. The mean velocity gradient corresponds to a depth of $0.09\pm0.05\,\rm{pc}$ along the line of sight, indicating that the analysis region consists of thin, filamentary structures as reported in \citetalias{2019ApJ...886...14F} and also seen in simulations \citep{2016MNRAS.457..375F,2018PASJ...70S..53I,2021ApJ...916...83A,2021NatAs...5..365F}. Further, we also find that the best-fit mean optical depths for the $^{12}$CO and $^{13}$CO lines are $17\pm4$ and $0.81\pm0.31$, and the excitation temperatures are $50\pm10\,\rm{K}$ and $39\pm6\,\rm{K}$, respectively. Thus, we confirm our hypothesis that $^{13}$CO is subthermally excited in the region. We see from \autoref{fig:columndensitymap} that the LTE analysis overpredicts the column density by $25-60$ per cent in regions where $^{13}$CO is sub-thermally excited, consistent with the recent findings of \cite{2021arXiv210611973F} based on a similar analysis in the LMC using the NLTE software library \texttt{RADEX} \citep{2007A&A...468..627V}. We next describe the procedure to extract the column density PDF in \autoref{s:hopkinsPDF}.

\begin{table}
\centering
\caption{Summary of the main parameters we derive to measure turbulent driving in the analysis region shown in \autoref{fig:fullmap}.}
\begin{tabular}{|l|l|l|l|}
\hline
Parameter & Description & LTE & NLTE\\
\hline
$T_{\rm{g}}/\rm{K}$ & Gas kinetic & $40\pm10$ & $58\pm8$\\
&temperature &&\\
$T_{\rm{ex}}/\rm{K}$ & $^{13}$CO excitation & $40\pm10$ & $39\pm6$\\
&temperature &&\\
$\log_{10}\,N_{0}/\rm{cm^{-2}}$ & Mean gas  & $22.8\pm0.05$ & $22.6\pm0.04$ \\
& column density &&\\
$\sigma_{N/N_0}$ & Column density & $0.58\pm0.14$ & $0.51\pm0.04$\\
&dispersion&&\\
$\theta$ & Column density & $0.12 \pm 0.05$ & $0.17 \pm 0.09$\\
&intermittency&&\\
$\sqrt{\mathcal{R}}$ & Degree of & $0.20 \pm 0.04$ & $0.15 \pm 0.03$\\
& anisotropy &&\\
$\sigma_{\rho/\rho_0}$ & 3D gas density  & $2.97^{+1.50}_{-0.71}$ & $3.40^{+0.92}_{-0.63}$\\
&dispersion&&\\
$\sigma_{\rm{v,1D}}/\rm{km\,s^{-1}}$ & 1D gas velocity & \multicolumn{2}{c}{$0.94\pm0.09$}\\
&dispersion&&\\
$c_{\rm{s}}/\rm{km\,s^{-1}}$ & Sound speed & $0.34\pm0.04$ & $0.41\pm0.03$\\
$\mathcal{M}$ & Turbulent Mach & $4.74^{+0.74}_{-0.54}$ & $3.92^{+0.35}_{-0.29}$\\
& number &&\\
\hline
$b$ & Turbulence driving & $0.62^{+0.34}_{-0.17}$ & $0.87^{+0.25}_{-0.17}$\\
& mode &&\\
\hline
\end{tabular}
\label{tab:tab1}
\end{table}

\subsubsection{Model for gas density PDF}
\label{s:hopkinsPDF}
We use our LTE and NLTE column density maps to create corresponding PDFs of $\eta = \log \left( N/N_0 \right)$ -- the natural logarithm of the column density $N$ normalized by the mean column density $N_0$. We estimate the width of each PDF $\sigma_\eta$ by fitting a \citet{Hopkins_2013} intermittent density PDF model to the volume-weighted PDF of $\eta$. This is a physically motivated fitting function that takes into account intermittent features in the density PDF, and provides a more accurate description of the density structure than the more common lognormal approximation, even for large turbulent Mach numbers, non-isothermal equations of state, and magnetic fields \citep{Hopkins_2013,Federrath_2015,Beattie2021b}.

However, we note that the \citet{Hopkins_2013} PDF model was originally developed for density PDFs and not column density PDFs. Here, we will assume that the column density PDF follows the same functional form as the density PDF. This is a reasonable assumption because simulations find that supersonic magnetohydrodynamic (MHD) turbulence results in very similar morphological features in the column and volume density PDFs, both of which are largely lognormal with some deviations from perfect Gaussianity (see figures~1 and 13 of \citealt{2007ApJ...658..423K}, and figure~1 of \citealt{2020ApJ...903L...2J}); these authors also show how the higher-order moments of the column and volume density scale with each other, which suggests that the same underlying PDF can be used for both quantities (see figure~14 of \citealt{2007ApJ...658..423K}). Furthermore, \citet{2009ApJ...693..250B} find that in MHD turbulence simulations, the correlations of several physical properties with density are similar to those with column density, thus justifying the use of the \cite{Hopkins_2013} model to derive the column density PDF parameters.

The Hopkins fitting function is given by
\begin{multline}
\label{eq:Hopkins}
    p_{\mathrm{HK}}(\eta)\,d\eta = I_1 \left(2 \sqrt{\lambda \omega(\eta)}\right) \exp {\left[ -\left(\lambda + \omega(\eta)\right) \right]} \sqrt{\frac{\lambda}{\theta^2 \omega(\eta)}}\,d\eta, \\
    \lambda = \frac{\sigma_{\eta}^2}{2 \theta^2}, \quad \omega(\eta) = \lambda/(1+\theta) - \eta/\theta \;\; (\omega \geq 0),
\end{multline}
where $I_1(x)$ is the first-order modified Bessel Function of the first kind, $\sigma_{\eta}$ is the standard deviation in $\eta$, and $\theta$ is the intermittency parameter that encapsulates the intermittent density PDF features. Note that in the zero-intermittency limit ($\theta \to 0$), \autoref{eq:Hopkins} simplifies to the lognormal PDF. We compute the best-fit parameters of $\sigma_\eta$ and $\theta$ using a bootstrapping approach with 10,000 realisations, where we create different random realisations based on Gaussian propagation of uncertainties in all the underlying parameters. We then transform $\sigma_\eta$ to the linear dispersion $\sigma_{N/N_0}$ by using the relation \citep{Hopkins_2013},
\begin{equation}
\label{eq:sigma_N_eta_relation}
    \sigma_{N/N_0} = \sqrt{\exp \left( {\frac{\sigma_\eta^2}{1+3\theta +2 \theta^2}} \right) -1} ,
\end{equation}
which can be derived from the moments of \autoref{eq:Hopkins}. We show the resulting PDFs and fits with the \cite{Hopkins_2013} model in \autoref{fig:coldumndensitypdf}. The shape of the two PDFs is very similar when normalized by $N_0$, as is also confirmed from their respective widths: $\sigma_{N/N_0} = 0.58 \pm 0.14$ and $0.51 \pm 0.04$ for the LTE and NLTE versions, respectively. We also find that the best-fit intermittency $\theta \sim 0.1-0.2$ in both the cases, reflecting the presence of non-lognormal density features in the region. The higher intermittency in the NLTE case is reflected in the fit slightly overshooting the peak in \autoref{fig:coldumndensitypdf}, but is consistent with the error margin in the data. Below, we will see that the turbulent Mach number we obtain is in good agreement with that expected from the intermittency-Mach number relation proposed by \cite{Hopkins_2013}.

There are several other models that characterise the non-lognormalities of the gas density PDF as a function of the supersonic turbulent Mach number \citep[e.g.,][]{2014A&A...565A..24F,2016MNRAS.460.4483K,2017MNRAS.471.3753S,Robertson2018,2018ApJ...865L..14S,2020ApJ...903L...2J}. In \aref{s:app_moczburkhart}, we fit one such PDF model given by \cite{Mocz2019}, which is an alternative intermittency density PDF model to estimate $\sigma_{N/N_0}$. We find that the results are consistent within the error bars. In addition to these, there are several models that use a lognormal function with a powerlaw function at higher densities to model the gas density PDF, where the powerlaw part is used to account for dense, self-gravitating structures in molecular clouds \citep[e.g.,][]{2011ApJ...727L..20K,2012ApJ...750...13C,2013ApJ...763...51F,2014ApJ...781...91G,2015ApJ...808...48B,2017ApJ...834L...1B,2018ApJ...863..118B,2019ApJ...879..129B,2020ApJ...903L...2J}. We also attempt to fit one such lognormal + powerlaw model from \cite{2021MNRAS.tmp.1659K}, finding that a purely lognormal fit is statistically preferred over a lognormal+powerlaw fit. Thus, we find that the \cite{Hopkins_2013} model provides the best fit to the data.

\begin{figure*}
\includegraphics[width=0.33\linewidth]{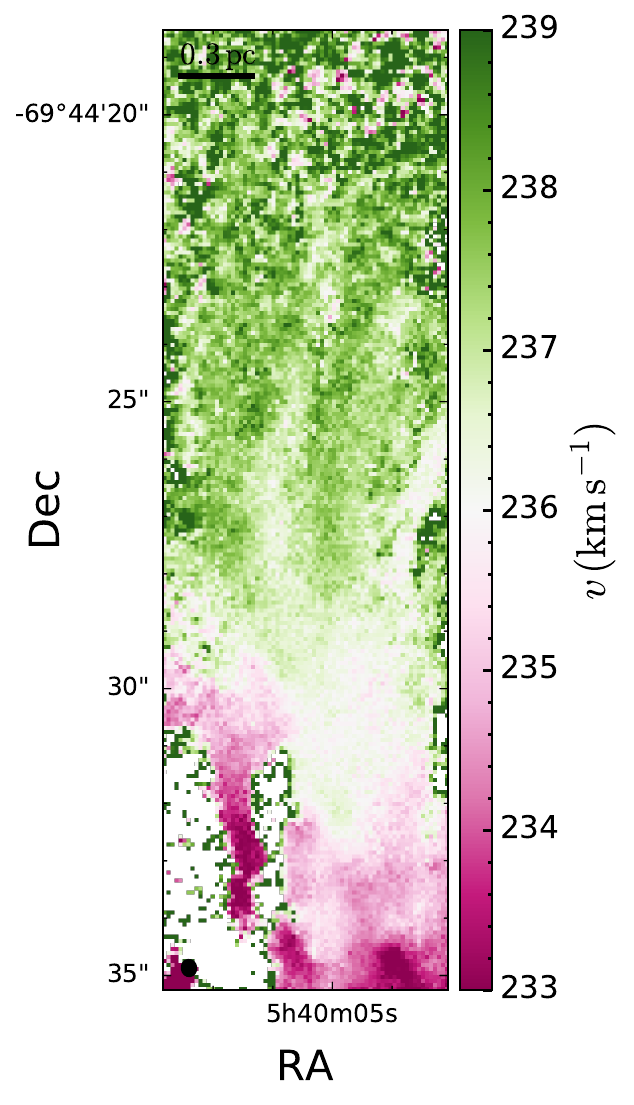}
\includegraphics[width=0.33\linewidth]{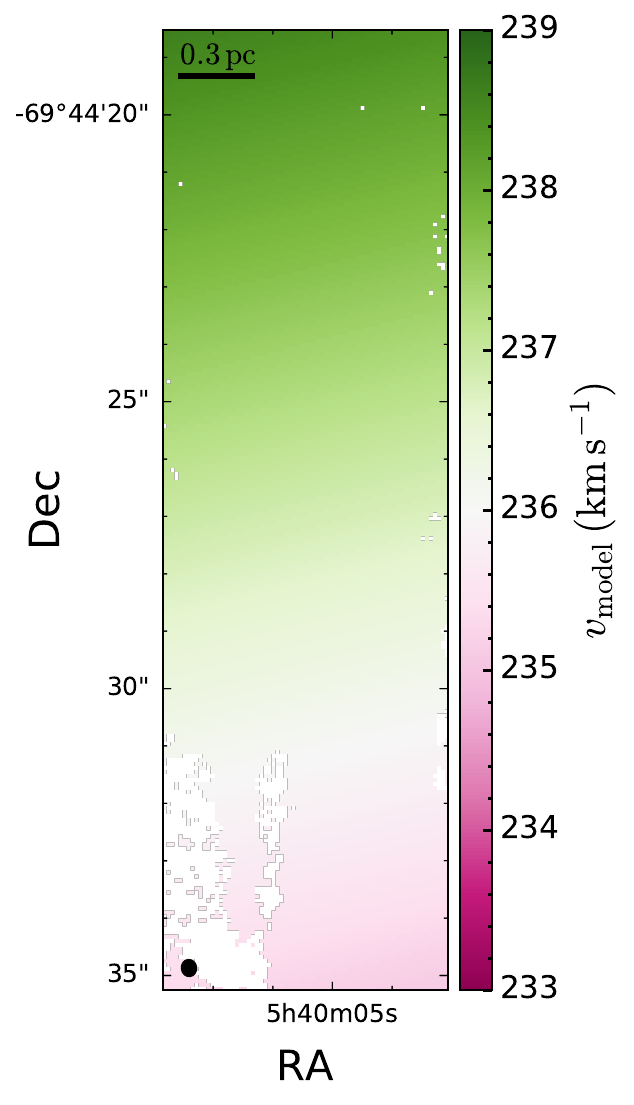}
\includegraphics[width=0.33\linewidth]{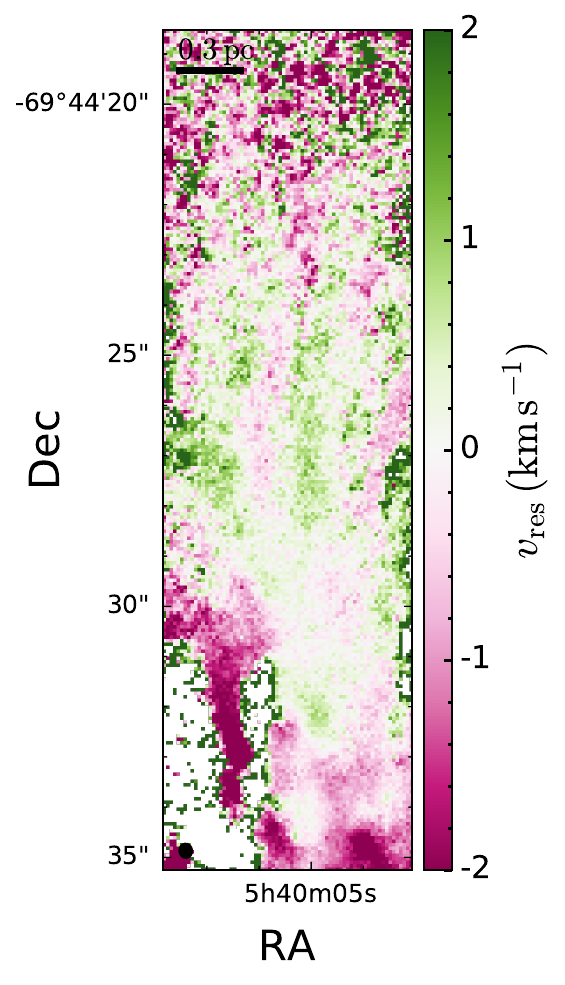}
\caption{The left panel shows the $^{13}$CO moment-1 (intensity-weighted velocity) map of the analysis region. The velocity map shows a large-scale gradient from top to bottom that we subtract by modeling it with a gradient as shown in the middle panel. The right panel shows the residual velocities after the large-scale gradient subtraction, which represent random, turbulent motions within the gas. We use the residual velocities together with the second moment map to find the overall 1D velocity dispersion in the gas, $\sigma_{\rm{v,1D}}$.}
\label{fig:vmaps}
\end{figure*}

\begin{figure*}
\includegraphics[width=1.0\columnwidth]{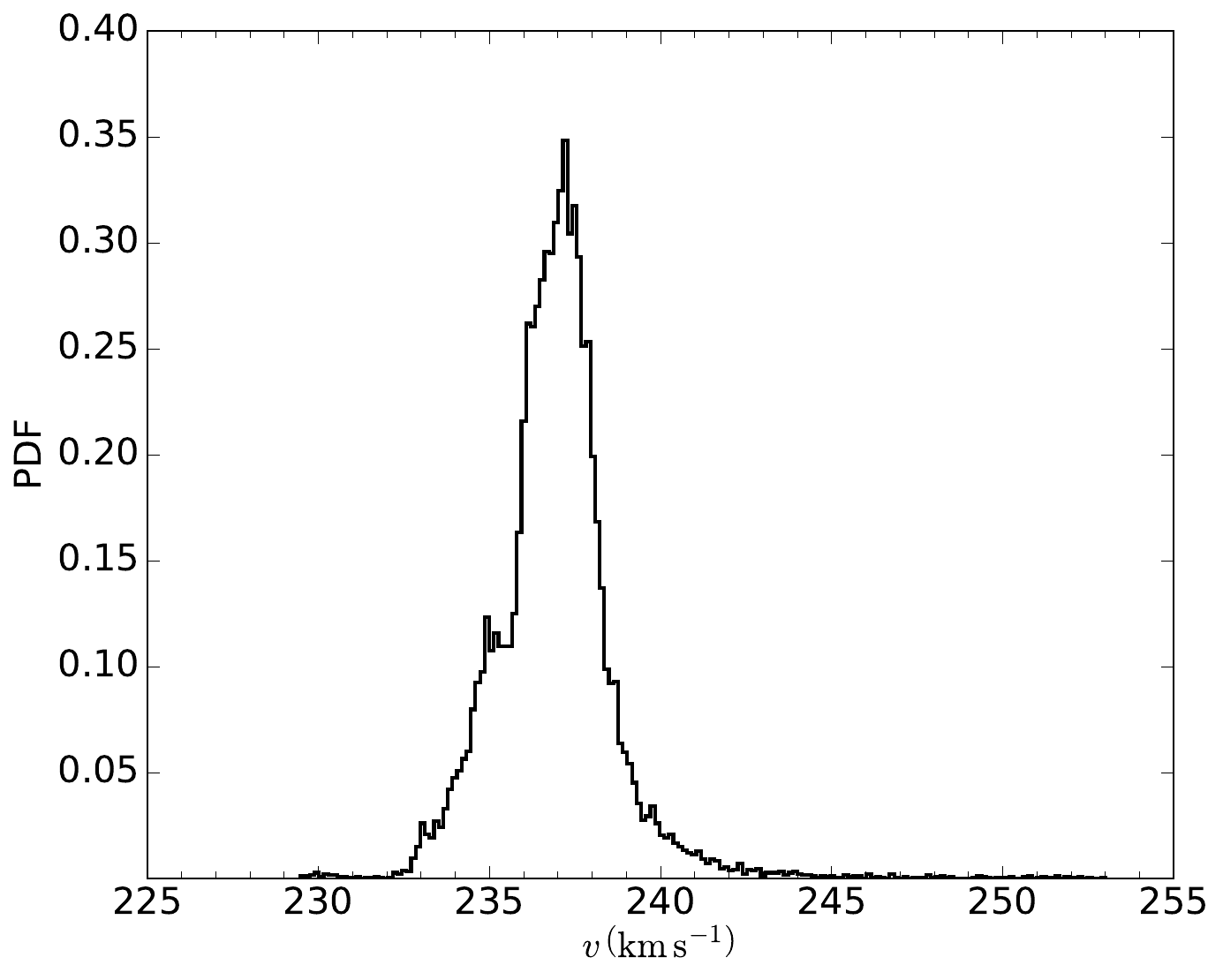}
\includegraphics[width=0.95\columnwidth]{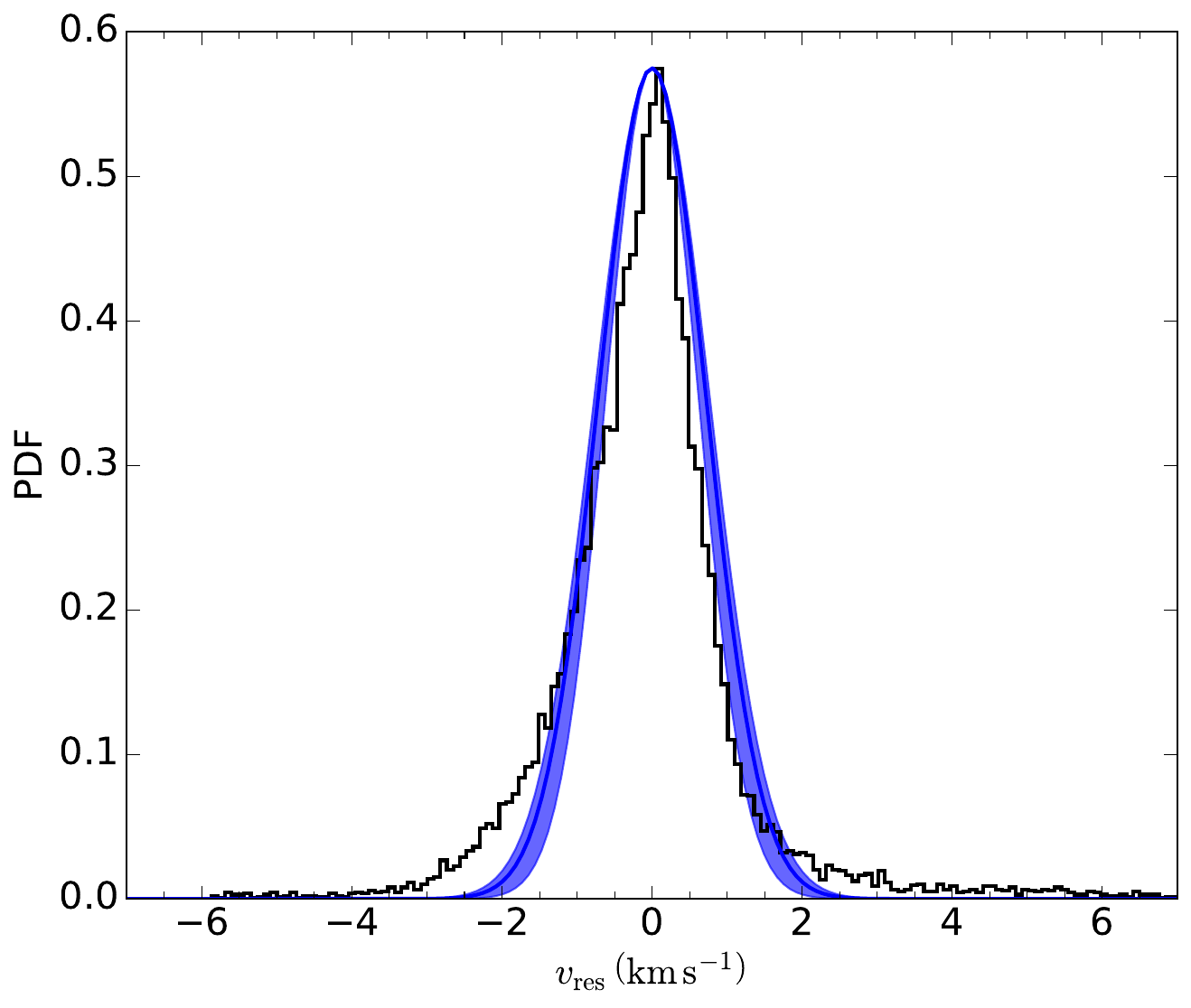}
\caption{PDFs of the velocity before (left) and after (right) gradient subtraction. The PDF of the residuals post gradient subtraction is largely Gaussian. The blue-shaded curve denotes the $1\sigma$ range of the best-fit Gaussian in solid blue. The width of the best-fit Gaussian (denoted by $\sigma_{\mathrm{v,1D,res}}$) together with the second moment velocity dispersion (denoted by $\sigma_{\rm{v,1D,2m}}$) is used to determine the overall 1D velocity dispersion in the gas, $\sigma_{\mathrm{v,1D}}$ (see \autoref{tab:tab1}). We find $\sigma_{\mathrm{v,1D,res}}=(0.72\pm0.05)\,\mathrm{km\,s^{-1}}$, and $\sigma_{\rm{v,1D}} = (0.94\pm0.09)\,\rm{km\,s^{-1}}$.}
\label{fig:vpdfs}
\end{figure*}

\subsubsection{Conversion from 2D to 3D density dispersion}
\label{sec:b_method_brunts_method}
To estimate the 3D density dispersion ($\sigma_{\rho/\rho_0}$) from the 2D column density dispersion ($\sigma_{N/N_0}$), we first measure the 2D column density power spectrum $P_\mathrm{2D}(k)$ of the quantity $N/N_0 - 1$, where $k = |\mathbf{k}|$ is the isotropic wavenumber. We use this to reconstruct the 3D density power spectrum $P_\mathrm{3D}(k)$ of the variable $\rho/\rho_0 -1$, as $P_\mathrm{3D}(k) = 2kP_\mathrm{2D}(k)$, assuming isotropy\footnote{This means that we assume the power in the spectral density follows the simple relation $P_\mathrm{3D}(k)/4\pi k^2 = P_\mathrm{2D}(k)/2\pi k \implies P_\mathrm{3D}(k) = 2kP_\mathrm{2D}(k)$, as stated above.} of the cloud \citep{Brunt_2010b,2010MNRAS.405L..56B,Kainulainen_2014}. The variance of a mean-zero field is proportional to the integral of the power spectrum over all $k$ \citep{Arfken_2013,Beattie2020a}. This implies that the 2D column density dispersion, $\sigma_{N/N_0}$ and 3D density dispersion, $\sigma_{\rho/\rho_0}$ are equal to $\left[\int P_\mathrm{2D}(k) dk\right]^{1/2}$ and $\left[\int P_\mathrm{3D}(k) dk\right]^{1/2}$, respectively. If we define $\mathcal{R} = P_\mathrm{2D}(k)/P_\mathrm{3D}(k)$, $\sigma_{\rho/\rho_0}$ can be obtained from the relation
\begin{equation}
\label{eq:Brunt_Equation}
   \sigma_{\rho/\rho_0} = \frac{\sigma_{N/N_0}}{\mathcal{\sqrt{R}}}\,.
\end{equation}
Thus, $\mathcal{R}$ represents the degree of anisotropy at the scale at which we measure $b$. \citet{Brunt_2010b} showed that \autoref{eq:Brunt_Equation} holds to within 10 per cent for isotropic, periodic fields, and is less accurate for non-periodic fields. We thus mimic a periodic dataset \citep{Ossenkopf_2008} by placing three mirrored copies of the column density map around itself in a periodic configuration. In the case of non-isotropic, periodic clouds, (for example, in the presence of anisotropic turbulence driving -- \citealt{2011ApJ...738...88H}, or strong mean magnetic fields -- \citealt{Beattie2020a}), the maximum uncertainty in the 2D-to-3D reconstruction for these cases is less than 40 per cent \citep[section~3.1.2]{2016ApJ...832..143F}, which is mostly noticed at lower turbulent Mach numbers than that we find below in \autoref{s:turbMach}. We use this upper limit as a systematic uncertainty in the 2D-to-3D reconstruction step in our analysis. Following this approach, we obtain a value of $\sqrt{\mathcal{R}} = 0.20 \pm 0.04$ and $0.15\pm0.03$, thus $\sigma_{\rho/\rho_0} = 2.97^{+1.50}_{-0.71}$ and $3.40^{+0.92}_{-0.63}$ from \autoref{eq:Brunt_Equation}, for the LTE and the NLTE analyses, respectively. The first panels in each row of \autoref{fig:distributions} show the PDF of $\sigma_{\rho/\rho_0}$ over all the bootstrapping realisations for the two cases, with the dashed lines depicting the $16^{\rm{th}}$, $50^{\rm{th}}$, and $84^{\rm{th}}$ percentiles, respectively. We see that the NLTE analysis gives a slightly higher 3D density dispersion than the LTE analysis, but it is consistent within the errorbars.

\subsection{Turbulent Mach number}
\label{s:turbMach}
The turbulent Mach number, $\mathcal{M}$, is defined as the ratio of the turbulent velocity dispersion in 3D, $\sigma_{\rm{v,3D}}$, to the sound speed, $c_{\rm{s}}$. The intensity-weighted velocity (moment 1) map ($v$) of $^{13}$CO of the analysis region reveals the presence of a systematic large-scale gradient in the gas, as we show in the left panel of \autoref{fig:vmaps}. To retrieve the turbulent motions within the gas, we fit a linear model ($v_{\rm{model}}$) to account for this large-scale systematic gradient and subtract it from the moment 1 map; the model is plotted in the middle panel of \autoref{fig:vmaps}. We then use the resulting residual map ($v_{\rm{res}}$), as shown in the right panel of \autoref{fig:vmaps}, to obtain the 1D velocity dispersion ($\sigma_{\rm{v,1D,res}}$) by modeling the PDF of the residual velocities as a Gaussian. \autoref{fig:vpdfs} shows the velocity PDFs before and after the subtraction of the large-scale gradient. Fitting the latter PDF with a Gaussian and using the resulting width as an estimate of $\sigma_{\rm{v,1D,res}}$, we obtain $\sigma_{\rm{v,1D,res}} = (0.72\pm 0.05)\,\rm{km\,s^{-1}}$. This technique has been used in several works (e.g., \citealt{2016ApJ...832..143F,2018ApJ...859..162C,2018MNRAS.477.4380S,2019MNRAS.487.4305S}, \citetalias{2021MNRAS.500.1721M}), and has been shown to be robust for deriving $\sigma_{\rm{v,1D,res}}$ provided the region is resolved with $\gtrsim 30$ resolution elements \citep{2018MNRAS.477.4380S}.\footnote{We do not measure turbulence driving in other regions in N159E (for example, the pillars identified by \citetalias{2019ApJ...886...14F} possibly created due to stellar radiation feedback around the Papillon Nebula) because we cannot derive a reasonable $\sigma_{\rm{v,1D,res}}$ for them using this technique.} 

However, there is no guarantee that residuals obtained from only removing the large-scale gradient represent the true random, turbulent motions within the cloud, since small-scale systematic motions or higher order moments can still persist and contribute to $\sigma_{\rm{v,1D,res}}$. In fact, we can see from the first and third panels of \autoref{fig:vmaps} that gradient subtraction has a limited effect on the southern part of the analysis region, which is also reflected in the negative velocity tail in the PDF of $v_{\rm{res}}$ in the right panel of \autoref{fig:vpdfs}. Recent work from \cite{2021stewart} finds that the above approach likely underestimates the true velocity dispersion, and a combination of $\sigma_{\rm{v,1D,res}}$ and the second moment (intensity-weighted velocity dispersion around the mean) gives a better estimate of the true velocity dispersion. So, we define the overall 1D velocity dispersion as $\sigma_{\rm{v,1D}} = \sqrt{(\sigma_{\rm{v,1D,res}})^2 + (\sigma_{\rm{v,1D,2m}})^2}$, where $\sigma_{\rm{v,1D,2m}} \approx 0.6\,\rm{km\,s^{-1}}$ is the velocity dispersion from the mean of the second moment of $^{13}$CO \citepalias{2019ApJ...886...14F}. Thus, we obtain $\sigma_{\rm{v,1D}} = (0.94\pm0.09)\,\rm{km\,s^{-1}}$. We then scale $\sigma_{\rm{v,1D}}$ by a factor $\sqrt{3}$ (assuming isotropic velocity fluctuations) to obtain the 3D velocity dispersion, $\sigma_{\rm{v,3D}}$. Using a combination of the gradient-subtracted first moment velocities and the second moment velocities to obtain the velocity dispersion rather than deriving it from the line shape in each pixel enables us to avoid the various biases that are introduced in line profiles; for example, flattened line centers due to opacity, overestimation of the turbulent Mach number due to opacity broadening, loss of line wings due to noise, or missing emission from low-density material due to excitation effects -- \citealt{2014ApJ...785L...1C,2016A&A...591A.104H,2020MNRAS.498.2440Y}.

For the LTE case, the turbulent Mach number corresponding to $T_{\rm{g}}=(40\pm10)\,\rm{K}$ is $\mathcal{M} = 4.74^{+0.74}_{-0.54}$. For the NLTE case, we find $\mathcal{M} = 3.92^{+0.35}_{-0.29}$ based on the best-fit $T_{\rm{g}}=(58\pm8)\,\rm{K}$. The turbulent Mach numbers we obtain are indicative of supersonic turbulence in the region and closely follow the $\mathcal{M}-\theta$ relation from \citet[figure~3]{Hopkins_2013}. The middle panel in each row of \autoref{fig:distributions} depicts the distribution of $\mathcal{M}$ for the two cases based on 10,000 bootstrapping realisations; the dashed lines denote the quantiles for the $16^{\rm{th}}$, $50^{\rm{th}}$, and $84^{\rm{th}}$ percentiles, respectively.

\begin{figure*}
\includegraphics[width=1.0\linewidth]{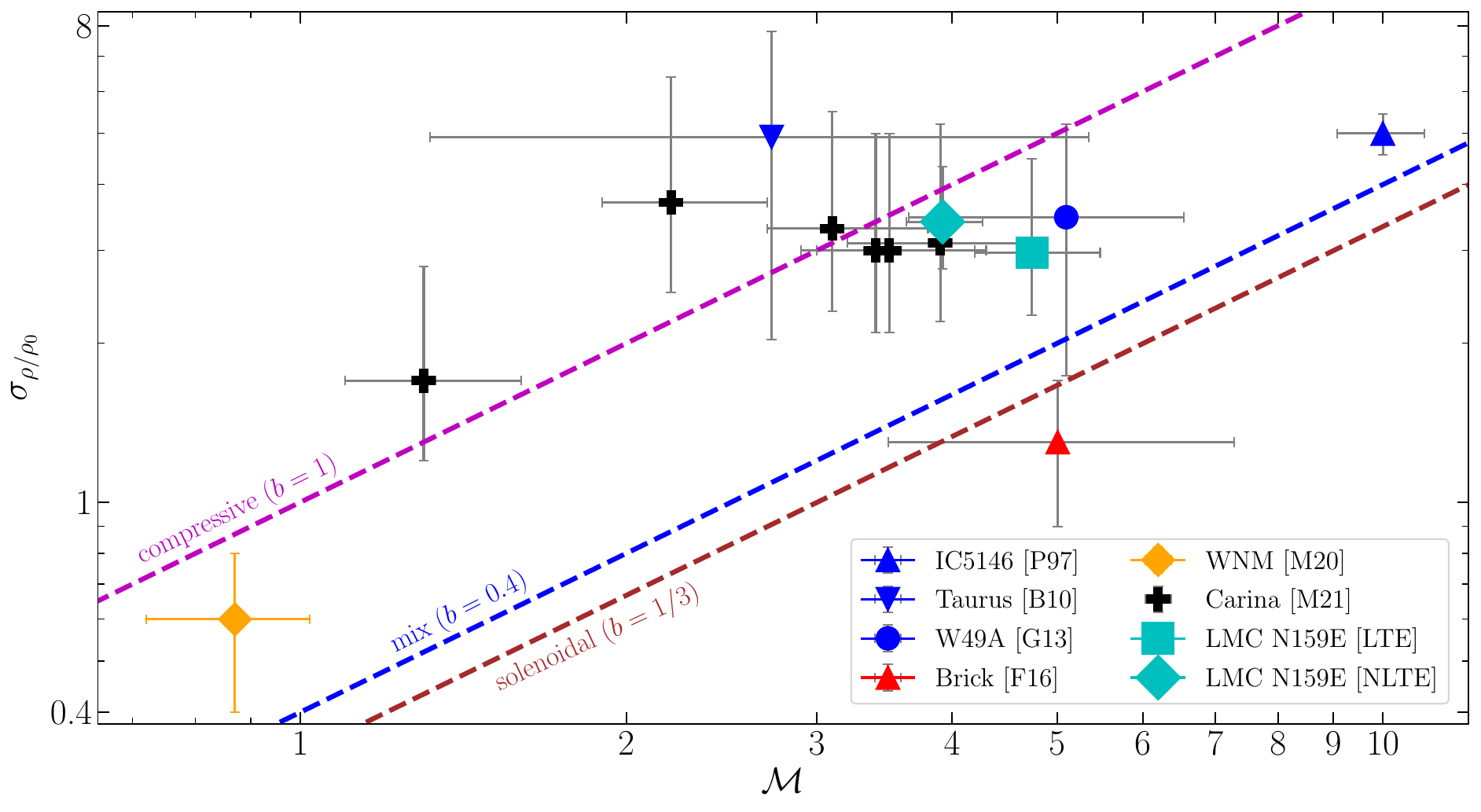}
\caption{Plot summarizing currently available measurements of the mode of turbulence from observations, denoted by the ratio of the 3D density dispersion ($\sigma_{\rho/\rho_0}$) to the turbulent Mach number ($\mathcal{M}$). The three dashed lines mark the theoretical limits within which turbulence can be driven through solenoidal or compressive modes, or a mixture of the two \protect\citep{2008ApJ...688L..79F,2010A&A...512A..81F}. The data is taken from various sources -- IC5146 \protect\citep{1997MNRAS.288..145P}, Taurus (\protect\citealt{2010A&A...513A..67B}, with revised estimates of the turbulent Mach number from \protect\citealt{2013A&A...549A..53K}, and magnetic field estimate from \protect\citealt{2016ApJ...832..143F}), W49A \protect\citep{2013ApJ...779...50G}, CMZ cloud Brick \protect\citep{2016ApJ...832..143F}, Warm Neutral Medium \protect\citep[WNM,][]{2020arXiv201203160M}, and 6 pillar-like regions in the Carina Nebula \protect\citepalias{2021MNRAS.500.1721M}. Errors are shown where available. Note that only Taurus and Brick have magnetic field estimates available for them, so for these clouds the $x$-axis should be read as $\mathcal{M}(1 + 1/\beta)^{-1/2}$, where $\beta$ is the turbulent plasma beta that accounts for the effects of magnetic fields in deriving the mode of turbulence \protect\citep{2012MNRAS.423.2680M}. The cyan markers denote the measurements for the LMC star-forming region N159E we study in this work using the LTE and the NLTE approaches; it is the only extragalactic region where the driving mode of turbulence has been measured.}
\label{fig:final}
\end{figure*}

\section{Results}
\label{s:resultsdiscussions}
Thanks to the exquisite resolution provided by ALMA, we can now derive the first extragalactic measurement of the driving mode of turbulence in a star-forming region. For simplicity, we collect all the different parameters we derive in \autoref{tab:tab1}. Using the measured values of the width of gas volume density PDF, $\sigma_{\rho/\rho_0}$, and turbulent Mach number, $\mathcal{M}$, we can now estimate the driving mode parameter $b = 0.62^{+0.34}_{-0.17}$ and $0.87^{+0.25}_{-0.17}$ for the LTE and the NLTE analyses, respectively. The right panels of the two rows in \autoref{fig:distributions} plots the distributions of $b$ resulting from the 10,000 bootstrapping realisations. These values of $b$ indicate the presence of primarily compressive (or, solenoidal-free) turbulence in the region. We also find that the more accurate NLTE analysis gives a higher $b$ as compared to the LTE analysis. Our findings (from the NLTE analysis) exclude a natural mix of modes and fully solenoidal driving at a $> 6\sigma$ and $>7\sigma$ level, respectively. 

It is worth noting that the distribution of $b$ in \autoref{fig:distributions} has a tail extending to $b>1$, which might seem in contradiction with theoretical expectations. While a part of the presence of this tail is simply due to observational and modeling uncertainties on various parameters, it could also be due to physical reasons. For example, it is possible that $b > 1$ if a physical process other than turbulence (such as gravity) has substantial contributions to density fluctuations (and hence the density dispersion). It is also possible that the derived turbulent Mach number includes contributions from non-turbulent sources, such as gravitational collapse motions. \citetalias{2021MNRAS.500.1721M} find that all of the pillars in the Carina Nebula where they derive $b>1$ have virial parameters < 1, suggesting that gravity plays a significant role in the density and velocity structure of gas in such pillars. On the other hand, \cite{2016ApJ...832..143F} find $b < 1/3$ for the Central Molecular Zone (CMZ) cloud Brick (also known as G0.253+0.016) due to strong shearing motions. Given the presence of dense cores like MMS-1 and MMS-2 in the analysis region, it is not surprising that the expected distribution of $b$ extends beyond unity. However, like our results, the uncertainties on the measurements of \cite{2016ApJ...832..143F} and \citetalias{2021MNRAS.500.1721M} are consistent with $1/3 \leq b \leq 1$.

It is helpful to place our results in the context of the literature. \autoref{fig:final} shows a compilation of observations where the driving mode of turbulence has been measured in the Galaxy. The data are limited and heterogeneous because different studies use different gas density tracers. Most clouds (except for Taurus and Brick) lack an estimate of the magnetic field strength, thus we assume $\beta \to \infty$ for them. For Taurus, we use the magnetic field estimate from \cite{2016ApJ...832..143F} that is based on earlier results from \cite{2012MNRAS.420.1562H}. We see two distinct classes of star-forming regions from \autoref{fig:final} -- one where the turbulence is highly compressive in nature (like the spiral arm molecular cloud Taurus and the star-forming pillars in the Carina Nebula), and another where it is largely non-compressive (like the CMZ cloud Brick). Our NLTE measurements for N159E (the only extragalactic region on this plot) fall in the former category. 

\section{Caveats}
\label{s:caveats}
The use of CO isotopologues to derive gas density PDFs typically introduces three major sources of uncertainty due to uncertain abundances, excitation temperatures, and fraction of CO-dark molecular gas. To test the impacts of abundance variations, we increase $X[\rm{H_2}/^{12}\mathrm{CO}]$ by a factor of 2 \citep[e.g.,][]{2013ARA&A..51..207B}, finding that it slightly decreases $\sigma_{\rho/\rho_0}$ without significantly changing the turbulence driving parameter $b$. Similarly, if we increase $X[^{12}\mathrm{CO}/^{13}\mathrm{CO}]$ to 70 \citep[typical of molecular clouds in the Galaxy --][]{1994A&A...291...89J,1994ARA&A..32..191W,2019ApJ...877..154Y}, $\sigma_{\rho/\rho_0}$ decreases by at most 17 per cent, which is within the error bars. The reason our results are insensitive to variations in CO abundances is because we use a combination of optically-thick and optically-thin tracers in our analysis (see \autoref{eq:chi2}). Additionally, the NLTE analysis ensures that we self-consistently calculate the optical depth of the CO isotopologues, thus ensuring that the radiative transfer effects do not bias the resulting density dispersion \citep{2013ApJ...771..122B}. While we have to assume $T_{\rm{g}}=T_{\rm{ex}}$ in LTE, the NLTE modeling gives a self-consistent excitation temperature, thus suppressing the uncertainties in $T_{\rm{ex}}$. We cannot constrain the amount of CO-dark molecular gas \citep[e.g.,][]{2012MNRAS.426..377G}, which can be quite high in low-metallicity regions like the LMC and the Small Magellanic Cloud \citep[SMC; e.g.,][]{2014ApJ...797...86R,2018ApJ...853..111J,2020MNRAS.494.5279C,2021arXiv210809018T}. Nonetheless, given the detection of C$^{18}$O in N159E, we do not expect to significantly underestimate high-density molecular gas within the region of interest. Thus, we find that the NLTE analysis mitigates some of the major caveats around assuming LTE. However, the LVG approximation used by the NLTE analysis can be problematic at very high CO column densities \citep{2018A&A...616A.131A}. This is because it only returns the integrated brightness temperatures for each line by assuming a top-hat line profile, thus ignoring the wings in the velocity-resolved line profiles we observe in the data. It also assumes a constant optical depth and escape probability for a given line at all frequencies. Future modeling of molecular gas tracers should therefore aim at providing velocity-resolved line profiles that are now measurable by ALMA.

There are also some caveats in extracting gas column density and constructing density PDFs using CO isotopologues. For example, the dynamic range of a density PDF constructed using CO isotopologues is quite narrow as compared to that probed by dust \citep[e.g.,][]{2013ApJ...766L..17S,2015A&A...578A..29S}. However, dust observations are typically at lower resolution than CO, and cannot be used to trace the gas kinematics \citep[e.g.,][]{2009ApJ...692...91G} or distinguish different clouds along the line of sight \citep[e.g.,][]{2016A&A...587A..74S}. It has also been shown that density PDFs constructed using interferometric observations can create a spurious low-density tail due to the direct current (DC) offset inherent to interferometry, and as a result can produce a PDF that is wider than the true PDF \citep{2016A&A...590A.104O}. Thus, it is quite possible that ALMA's interferometric data leads to an overestimation of $b$. We have also ignored the effects of magnetic fields throughout this analysis. If significant, magnetic fields would lead to an increase in $b$, thereby favouring a more compressive turbulent driving \citep{2011ApJ...730...40P,2012ApJ...761..156F,2012MNRAS.423.2680M}. For example, if we follow \cite{2012ARA&A..50...29C} to get a rough estimate of the magnetic field strength based on the mean column density in the region, we find a field strength $\sim 100\,\rm{\mu G}$. This field strength corresponds to turbulent plasma $\beta \sim 3.5$ for the mean density corresponding to the NLTE case, thus giving $b \sim 0.98$. Since magnetic fields and possible overestimation of $\sigma_{\rho/\rho_0}$ due to interferometric effects alter $b$ in opposite directions, we speculate that their effects on the driving mode roughly balance out. However, it is difficult to predict the exact changes in $b$ without an accurate estimate of the magnetic field strength. Future observations (for example, using Faraday rotation measure -- \citealt{2005Sci...307.1610G,2012ApJ...759...25M,2021MNRAS.502.3814L}) will provide critical constraints on the magnetic field strength in resolved regions within the Magellanic Clouds.

\section{Possible drivers of compressive turbulence in N159E}
\label{s:possibledrivers}
If the density structure and the dynamics we see in N159E is highly influenced by gravitational collapse motions, then we naturally expect a large fraction of compressive modes to be introduced, which may explain the high value of $b$ \citep{2020ApJ...903L...2J,2021MNRAS.tmp.1659K}. Several studies have shown how supersonic turbulence can compress the gas on local scales leading to the creation of cores (such as MMS-1 and MMS-2 in N159E) and filaments, while stabilizing the cloud against gravity on larger scales \citep[e.g.,][]{2002A&A...390..307O,2011A&A...529A...1S,2012ApJ...761..156F,Robertson2018,Beattie2020b,Imara2021}. To verify whether this is indeed the case, we require observations of molecular gas tracers for higher densities (e.g., HCN or HCO$^+$) at similar resolution and sensitivity as the CO observations we use in this study. This is because such tracers can potentially reveal the power-law tail in the density PDF due to self-gravitating structures, such as the MMS-1 and MMS-2 cores.

Another plausible candidate for compressive driving is the compression caused by the \hi flow that is proposed to have led to the creation of the massive molecular filaments observed by ALMA \citepalias[][figure~8]{2019ApJ...886...14F}, a view supported by magnetohydrodynamical simulations \citep{2012ApJ...759...35I,2013ApJ...774L..31I}. This flow is proposed to have originated due to tidal interactions between the LMC and SMC \citep{2007PASA...24...21B,2017PASJ...69L...5F} $0.2\,\rm{Gyr}$ ago \citep{1990PASJ...42..505F}. However, the limited resolution of currently available \hi data \citep{2003ApJS..148..473K} makes it challenging to study the correlation between the molecular gas structure observed by ALMA and the larger-scale \hi gas structure \citep[e.g.,][]{2009ApJ...705..144F}. Note that N159E is only one part of the GMC N159. The western end of N159 (denoted by N159W in the literature -- \citealt{1998A&A...332..493H}) shows similar molecular gas morphology, protostellar activity, and star formation activity \citep{2019ApJ...886...15T}. These similarities hint at a common origin of the kinematics and dynamics in both N159E and N159W, providing support for the large-scale \hi flow scenario \citepalias{2019ApJ...886...14F}. Even if N159W also exhibits largely compressive turbulence, it is difficult to establish whether the \hi flow that originated $0.2\,\rm{Gyr}$ ago is the cause, given the instantaneous nature of the mode of turbulence that we can measure from observations. High-resolution \hi observations from the \hi component of the upcoming Galactic ASKAP survey (GASKAP-\hi -- \citealt{2013PASA...30....3D,2021pingel}) and eventual SKA observations \citep{2020Ap&SS.365..118K} will prove instrumental in pinpointing the driver of compressive turbulence in N159E.

As we noted in \autoref{s:intro}, protostellar outflows can also drive and inject turbulence in star-forming regions. Simulations find that protostellar outflows drive primarily solenoidal motions, at least in the shearing layers between outflow and ambient gas \citep{2014ApJ...784...61O,2017ApJ...847..104O}, while the bow shock of an outflow is more compressive \citep[e.g.,][]{angeo-28-951-2010}. However, the protostellar outflows observed in N159E are only $\sim 0.1\,\rm{pc}$ in size \citepalias{2019ApJ...886...14F}, so we do not expect them to drive turbulence in our analysis region, which is $10\times$ bigger (see \autoref{fig:fullmap}). Thus, we can exclude them as a source of the compressive driving of turbulence in this region. 

Lastly, it is worth emphasizing that turbulence driving in other parts of N159E not covered in this work may not be compressive, or may not share the same physical origin for compressive driving as the analysis region. For example, it is likely that the CO pillars surrounding the Papillon YSO (magenta box in \autoref{fig:fullmap}) also exhibit compressive turbulence driving, but due to photoionization feedback from the YSOs, similar to the CO pillars in the Carina nebula in the Milky Way \citepalias{2021MNRAS.500.1721M}.

\section{Conclusions}
\label{s:conclusions}
In this work, we utilize high-resolution (sub-pc) ALMA $^{12}$CO ($J$~=~2--1), $^{13}$CO ($J$~=~2--1), and C$^{18}$O ($J$~=~2--1) observations of the star-forming region N159E (the Papillon Nebula) in the LMC to provide the first extragalactic measurement of turbulence driving at pc scales. The turbulence driving mode is characterized by the parameter $b$ -- the ratio of the normalized gas density dispersion ($\sigma_{\rho/\rho_0}$) to the turbulent Mach number ($\mathcal{M}$). Values of $b \approx 1/3$ represent a highly solenoidal driving, $b\approx 1$ represents highly compressive driving, and a natural mixture of the two is given by $b\approx0.4$ \citep{2008ApJ...688L..79F,2010A&A...512A..81F}. 

We use an NLTE analysis of the CO isotopologues using the radiative transfer code \texttt{DESPOTIC} \citep{2014MNRAS.437.1662K} to construct the gas column density PDF for the analysis region we show in \autoref{fig:fullmap}, which consists of filamentary molecular clouds with dense cores driving protostellar outflows. We then fit the PDF with the \cite{Hopkins_2013} density PDF model to get the 2D column density dispersion, which we then convert to get the 3D density dispersion using the \cite{Brunt_2010b} method. We obtain $\sigma_{\rho/\rho_0} = 3.4^{+0.92}_{-0.63}$. We use a combination of the gradient-subtracted first moment map and the mean of the second moment map to get the gas velocity dispersion, which gives $\mathcal{M} = 3.9^{+0.35}_{-0.29}$. Thus, the driving mode of turbulence we obtain is $b = 0.87^{+0.25}_{-0.17}$, indicating that the turbulence being driven in this region is highly compressive. We suspect that the compressive turbulence could be driven by gravo-turbulent fragmentation of the dense gas, or by the \hi flow that is proposed to have led to the creation of GMC N159. This analysis is easily reproducible and can be applied to measure $b$ in resolved star-forming regions both in the local and the high-redshift Universe.

\section*{Acknowledgements}
We thank Tony Wong for going through a preprint of this paper and providing comments. We also thank an anonymous referee for a constructive feedback on our work. We acknowledge useful discussions with Shivan Khullar on gas density PDFs, with Jack Livingston on magnetic fields in the LMC, and with Rajsekhar Mohapatra on the mode of turbulence. PS is supported by the Australian Government Research Training Program (RTP) Scholarship. CF and MRK acknowledge funding provided by the Australian Research Council (ARC) through Discovery Projects DP170100603 (CF) and DP190101258 (MRK) and Future Fellowships FT180100495 (CF) and FT180100375 (MRK), and by the Australia-Germany Joint Research Cooperation Scheme grant (UA-DAAD). MRK is also the recipient of an Alexander von Humboldt award. PS, CF and MRK acknowledge the support of the ARC Centre of Excellence for All Sky Astrophysics in 3 Dimensions (ASTRO 3D), through project number CE170100013. JRB acknowledges financial support from the Australian National University via the Deakin PhD and Dean’s Higher Degree Research (theoretical physics) Scholarships. KT is supported by the National Astronomical Observatory of Japan (NAOJ) ALMA Scientific Research Grant Number 2016-03B. BB acknowledges the generous support of the Flatiron Institute Simons Foundation and National Aeronautics and Space Administration (NASA) award 19-ATP19-0020. CJL acknowledges funding from the NSF Graduate Research Fellowship under grant DGE1745303. TJG acknowledges support under NASA contract NAS8-03060 with the \textit{Chandra} X-ray Center. NMP acknowledges from ARC DP190101571 and National Science Foundation (NSF) grant AST-1309815. IRS acknowledges support from the ARC grant FT160100028. HS is supported by Japan Society for the Promotion of Science (JSPS) KAKENHI Grant Numbers JP19H05075 and JP21H01136. This paper makes use of the following ALMA data: ADS/JAO. ALMA\#2016.1.01173.S. ALMA is a partnership of the ESO, NSF, NINS, NRC, NSC, and ASIAA. The Joint ALMA Observatory is operated by the ESO, AUI/NRAO, and NAOJ. Analysis was performed using \texttt{NUMPY} \citep{oliphant2006guide,2020arXiv200610256H}, \texttt{ASTROPY} \citep{2013A&A...558A..33A,2018AJ....156..123A} and \texttt{SCIPY} \citep{2020NatMe..17..261V}; plots were created using \texttt{MATPLOTLIB} \citep{Hunter:2007}. This research has made extensive use of LAMDA \citep{2005A&A...432..369S,2020Atoms...8...15V} and NASA's Astrophysics Data System Bibliographic Services. The LAMDA database is supported by the Netherlands Organization for Scientific Research (NWO), the Netherlands Research School for Astronomy (NOVA), and the Swedish Research Council.

%%%%%%%%%%%%%%%%%%%%%%%%%%%%%%%%%%%%%%%%%%%%%%%%%%
\section*{Data Availability Statement}
No new data were generated for this work. The ALMA data we use in this work is publicly available \hyperlink{https://almascience.nrao.edu/}{here}.



\bibliographystyle{mnras}
\bibliography{references}

\begin{thebibliography}{}
\makeatletter
\relax
\def\mn@urlcharsother{\let\do\@makeother \do\$\do\&\do\#\do\^\do\_\do\%\do\~}
\def\mn@doi{\begingroup\mn@urlcharsother \@ifnextchar [ {\mn@doi@}
  {\mn@doi@[]}}
\def\mn@doi@[#1]#2{\def\@tempa{#1}\ifx\@tempa\@empty \href
  {http://dx.doi.org/#2} {doi:#2}\else \href {http://dx.doi.org/#2} {#1}\fi
  \endgroup}
\def\mn@eprint#1#2{\mn@eprint@#1:#2::\@nil}
\def\mn@eprint@arXiv#1{\href {http://arxiv.org/abs/#1} {{\tt arXiv:#1}}}
\def\mn@eprint@dblp#1{\href {http://dblp.uni-trier.de/rec/bibtex/#1.xml}
  {dblp:#1}}
\def\mn@eprint@#1:#2:#3:#4\@nil{\def\@tempa {#1}\def\@tempb {#2}\def\@tempc
  {#3}\ifx \@tempc \@empty \let \@tempc \@tempb \let \@tempb \@tempa \fi \ifx
  \@tempb \@empty \def\@tempb {arXiv}\fi \@ifundefined
  {mn@eprint@\@tempb}{\@tempb:\@tempc}{\expandafter \expandafter \csname
  mn@eprint@\@tempb\endcsname \expandafter{\@tempc}}}

\bibitem[\protect\citeauthoryear{{Abe}, {Inoue}, {Inutsuka}  \&
  {Matsumoto}}{{Abe} et~al.}{2021}]{2021ApJ...916...83A}
{Abe} D.,  {Inoue} T.,  {Inutsuka} S.-i.,   {Matsumoto} T.,  2021, \mn@doi
  [\apj] {10.3847/1538-4357/ac07a1}, \href
  {https://ui.adsabs.harvard.edu/abs/2021ApJ...916...83A} {916, 83}

\bibitem[\protect\citeauthoryear{Arfken, Weber  \& Harris}{Arfken
  et~al.}{2013}]{Arfken_2013}
Arfken G.~B.,  Weber H.~J.,   Harris F.~E.,  2013, in Arfken G.~B.,  Weber
  H.~J.,   Harris F.~E.,  eds, , Mathematical Methods for Physicists (Seventh
  Edition), seventh edition edn, Academic Press, Boston, pp 935 -- 962

\bibitem[\protect\citeauthoryear{{Asensio Ramos} \& {Elitzur}}{{Asensio Ramos}
  \& {Elitzur}}{2018}]{2018A&A...616A.131A}
{Asensio Ramos} A.,  {Elitzur} M.,  2018, \mn@doi [\aap]
  {10.1051/0004-6361/201731943}, \href
  {https://ui.adsabs.harvard.edu/abs/2018A&A...616A.131A} {616, A131}

\bibitem[\protect\citeauthoryear{{Astropy Collaboration} et~al.,}{{Astropy
  Collaboration} et~al.}{2013}]{2013A&A...558A..33A}
{Astropy Collaboration} et~al., 2013, \mn@doi [\aap]
  {10.1051/0004-6361/201322068}, \href
  {https://ui.adsabs.harvard.edu/abs/2013A\%26A...558A..33A} {558, A33}

\bibitem[\protect\citeauthoryear{{Astropy Collaboration} et~al.,}{{Astropy
  Collaboration} et~al.}{2018}]{2018AJ....156..123A}
{Astropy Collaboration} et~al., 2018, \mn@doi [\aj] {10.3847/1538-3881/aabc4f},
  \href {https://ui.adsabs.harvard.edu/abs/2018AJ....156..123A} {156, 123}

\bibitem[\protect\citeauthoryear{{Beattie} \& {Federrath}}{{Beattie} \&
  {Federrath}}{2020}]{Beattie2020a}
{Beattie} J.~R.,  {Federrath} C.,  2020, \mn@doi [\mnras]
  {10.1093/mnras/stz3377}, \href
  {https://ui.adsabs.harvard.edu/abs/2020MNRAS.492..668B} {492, 668}

\bibitem[\protect\citeauthoryear{{Beattie}, {Federrath}  \& {Seta}}{{Beattie}
  et~al.}{2020}]{Beattie2020b}
{Beattie} J.~R.,  {Federrath} C.,   {Seta} A.,  2020, \mn@doi [\mnras]
  {10.1093/mnras/staa2257}, \href
  {https://ui.adsabs.harvard.edu/abs/2020MNRAS.498.1593B} {498, 1593}

\bibitem[\protect\citeauthoryear{{Beattie}, {Mocz}, {Federrath}  \&
  {Klessen}}{{Beattie} et~al.}{2021a}]{Beattie2021b}
{Beattie} J.~R.,  {Mocz} P.,  {Federrath} C.,   {Klessen} R.~S.,  2021a, arXiv
  e-prints, \href {https://ui.adsabs.harvard.edu/abs/2021arXiv210910470B} {p.
  arXiv:2109.10470}

\bibitem[\protect\citeauthoryear{{Beattie}, {Mocz}, {Federrath}  \&
  {Klessen}}{{Beattie} et~al.}{2021b}]{2021MNRAS.504.4354B}
{Beattie} J.~R.,  {Mocz} P.,  {Federrath} C.,   {Klessen} R.~S.,  2021b,
  \mn@doi [\mnras] {10.1093/mnras/stab1037}, \href
  {https://ui.adsabs.harvard.edu/abs/2021MNRAS.504.4354B} {504, 4354}

\bibitem[\protect\citeauthoryear{{Bekki} \& {Chiba}}{{Bekki} \&
  {Chiba}}{2007}]{2007PASA...24...21B}
{Bekki} K.,  {Chiba} M.,  2007, \mn@doi [\pasa] {10.1071/AS06023}, \href
  {https://ui.adsabs.harvard.edu/abs/2007PASA...24...21B} {24, 21}

\bibitem[\protect\citeauthoryear{{Bolatto}, {Wolfire}  \& {Leroy}}{{Bolatto}
  et~al.}{2013}]{2013ARA&A..51..207B}
{Bolatto} A.~D.,  {Wolfire} M.,   {Leroy} A.~K.,  2013, \mn@doi [\araa]
  {10.1146/annurev-astro-082812-140944}, \href
  {https://ui.adsabs.harvard.edu/abs/2013ARA&A..51..207B} {51, 207}

\bibitem[\protect\citeauthoryear{{Brunt}}{{Brunt}}{2010}]{2010A&A...513A..67B}
{Brunt} C.~M.,  2010, \mn@doi [\aap] {10.1051/0004-6361/200913506}, \href
  {https://ui.adsabs.harvard.edu/abs/2010A&A...513A..67B} {513, A67}

\bibitem[\protect\citeauthoryear{{Brunt}, {Federrath}  \& {Price}}{{Brunt}
  et~al.}{2010a}]{Brunt_2010b}
{Brunt} C.~M.,  {Federrath} C.,   {Price} D.~J.,  2010a, \mn@doi [\mnras]
  {10.1111/j.1365-2966.2009.16215.x}, \href
  {https://ui.adsabs.harvard.edu/abs/2010MNRAS.403.1507B} {403, 1507}

\bibitem[\protect\citeauthoryear{{Brunt}, {Federrath}  \& {Price}}{{Brunt}
  et~al.}{2010b}]{2010MNRAS.405L..56B}
{Brunt} C.~M.,  {Federrath} C.,   {Price} D.~J.,  2010b, \mn@doi [\mnras]
  {10.1111/j.1745-3933.2010.00858.x}, \href
  {https://ui.adsabs.harvard.edu/abs/2010MNRAS.405L..56B} {405, L56}

\bibitem[\protect\citeauthoryear{{Burkhart}}{{Burkhart}}{2018}]{2018ApJ...863..118B}
{Burkhart} B.,  2018, \mn@doi [\apj] {10.3847/1538-4357/aad002}, \href
  {https://ui.adsabs.harvard.edu/abs/2018ApJ...863..118B} {863, 118}

\bibitem[\protect\citeauthoryear{{Burkhart} \& {Lazarian}}{{Burkhart} \&
  {Lazarian}}{2012}]{2012ApJ...755L..19B}
{Burkhart} B.,  {Lazarian} A.,  2012, \mn@doi [\apjl]
  {10.1088/2041-8205/755/1/L19}, \href
  {https://ui.adsabs.harvard.edu/abs/2012ApJ...755L..19B} {755, L19}

\bibitem[\protect\citeauthoryear{{Burkhart} \& {Mocz}}{{Burkhart} \&
  {Mocz}}{2019}]{2019ApJ...879..129B}
{Burkhart} B.,  {Mocz} P.,  2019, \mn@doi [\apj] {10.3847/1538-4357/ab25ed},
  \href {https://ui.adsabs.harvard.edu/abs/2019ApJ...879..129B} {879, 129}

\bibitem[\protect\citeauthoryear{{Burkhart}, {Falceta-Gon{\c{c}}alves}, {Kowal}
   \& {Lazarian}}{{Burkhart} et~al.}{2009}]{2009ApJ...693..250B}
{Burkhart} B.,  {Falceta-Gon{\c{c}}alves} D.,  {Kowal} G.,   {Lazarian} A.,
  2009, \mn@doi [\apj] {10.1088/0004-637X/693/1/250}, \href
  {https://ui.adsabs.harvard.edu/abs/2009ApJ...693..250B} {693, 250}

\bibitem[\protect\citeauthoryear{{Burkhart}, {Ossenkopf}, {Lazarian}  \&
  {Stutzki}}{{Burkhart} et~al.}{2013}]{2013ApJ...771..122B}
{Burkhart} B.,  {Ossenkopf} V.,  {Lazarian} A.,   {Stutzki} J.,  2013, \mn@doi
  [\apj] {10.1088/0004-637X/771/2/122}, \href
  {https://ui.adsabs.harvard.edu/abs/2013ApJ...771..122B} {771, 122}

\bibitem[\protect\citeauthoryear{{Burkhart}, {Collins}  \&
  {Lazarian}}{{Burkhart} et~al.}{2015}]{2015ApJ...808...48B}
{Burkhart} B.,  {Collins} D.~C.,   {Lazarian} A.,  2015, \mn@doi [\apj]
  {10.1088/0004-637X/808/1/48}, \href
  {https://ui.adsabs.harvard.edu/abs/2015ApJ...808...48B} {808, 48}

\bibitem[\protect\citeauthoryear{{Burkhart}, {Stalpes}  \&
  {Collins}}{{Burkhart} et~al.}{2017}]{2017ApJ...834L...1B}
{Burkhart} B.,  {Stalpes} K.,   {Collins} D.~C.,  2017, \mn@doi [\apjl]
  {10.3847/2041-8213/834/1/L1}, \href
  {https://ui.adsabs.harvard.edu/abs/2017ApJ...834L...1B} {834, L1}

\bibitem[\protect\citeauthoryear{{Burkhart} et~al.,}{{Burkhart}
  et~al.}{2020}]{2020ApJ...905...14B}
{Burkhart} B.,  et~al., 2020, \mn@doi [\apj] {10.3847/1538-4357/abc484}, \href
  {https://ui.adsabs.harvard.edu/abs/2020ApJ...905...14B} {905, 14}

\bibitem[\protect\citeauthoryear{{Chen} et~al.,}{{Chen}
  et~al.}{2010}]{2010ApJ...721.1206C}
{Chen} C. H.~R.,  et~al., 2010, \mn@doi [\apj] {10.1088/0004-637X/721/2/1206},
  \href {https://ui.adsabs.harvard.edu/abs/2010ApJ...721.1206C} {721, 1206}

\bibitem[\protect\citeauthoryear{{Chen}, {Burkhart}, {Goodman}  \&
  {Collins}}{{Chen} et~al.}{2018}]{2018ApJ...859..162C}
{Chen} H. H.-H.,  {Burkhart} B.,  {Goodman} A.,   {Collins} D.~C.,  2018,
  \mn@doi [\apj] {10.3847/1538-4357/aabaf6}, \href
  {https://ui.adsabs.harvard.edu/abs/2018ApJ...859..162C} {859, 162}

\bibitem[\protect\citeauthoryear{{Chevance} et~al.,}{{Chevance}
  et~al.}{2020}]{2020MNRAS.494.5279C}
{Chevance} M.,  et~al., 2020, \mn@doi [\mnras] {10.1093/mnras/staa1106}, \href
  {https://ui.adsabs.harvard.edu/abs/2020MNRAS.494.5279C} {494, 5279}

\bibitem[\protect\citeauthoryear{{Choi}, {Evans}  \& {Jaffe}}{{Choi}
  et~al.}{1993}]{Choi_1993}
{Choi} M.,  {Evans} Neal~J. I.,   {Jaffe} D.~T.,  1993, \mn@doi [\apj]
  {10.1086/173341}, \href
  {https://ui.adsabs.harvard.edu/abs/1993ApJ...417..624C} {417, 624}

\bibitem[\protect\citeauthoryear{{Collins}, {Kritsuk}, {Padoan}, {Li}, {Xu},
  {Ustyugov}  \& {Norman}}{{Collins} et~al.}{2012}]{2012ApJ...750...13C}
{Collins} D.~C.,  {Kritsuk} A.~G.,  {Padoan} P.,  {Li} H.,  {Xu} H.,
  {Ustyugov} S.~D.,   {Norman} M.~L.,  2012, \mn@doi [\apj]
  {10.1088/0004-637X/750/1/13}, \href
  {https://ui.adsabs.harvard.edu/abs/2012ApJ...750...13C} {750, 13}

\bibitem[\protect\citeauthoryear{{Correia}, {Burkhart}, {Lazarian},
  {Ossenkopf}, {Stutzki}, {Kainulainen}, {Kowal}  \& {de Medeiros}}{{Correia}
  et~al.}{2014}]{2014ApJ...785L...1C}
{Correia} C.,  {Burkhart} B.,  {Lazarian} A.,  {Ossenkopf} V.,  {Stutzki} J.,
  {Kainulainen} J.,  {Kowal} G.,   {de Medeiros} J.~R.,  2014, \mn@doi [\apjl]
  {10.1088/2041-8205/785/1/L1}, \href
  {https://ui.adsabs.harvard.edu/abs/2014ApJ...785L...1C} {785, L1}

\bibitem[\protect\citeauthoryear{{Crutcher}}{{Crutcher}}{2012}]{2012ARA&A..50...29C}
{Crutcher} R.~M.,  2012, \mn@doi [\araa] {10.1146/annurev-astro-081811-125514},
  \href {https://ui.adsabs.harvard.edu/abs/2012ARA&A..50...29C} {50, 29}

\bibitem[\protect\citeauthoryear{{Dickey} et~al.,}{{Dickey}
  et~al.}{2013}]{2013PASA...30....3D}
{Dickey} J.~M.,  et~al., 2013, \mn@doi [\pasa] {10.1017/pasa.2012.003}, \href
  {https://ui.adsabs.harvard.edu/abs/2013PASA...30....3D} {30, e003}

\bibitem[\protect\citeauthoryear{{Federrath}}{{Federrath}}{2016}]{2016MNRAS.457..375F}
{Federrath} C.,  2016, \mn@doi [\mnras] {10.1093/mnras/stv2880}, \href
  {https://ui.adsabs.harvard.edu/abs/2016MNRAS.457..375F} {457, 375}

\bibitem[\protect\citeauthoryear{{Federrath} \& {Banerjee}}{{Federrath} \&
  {Banerjee}}{2015}]{Federrath_2015}
{Federrath} C.,  {Banerjee} S.,  2015, \mn@doi [\mnras] {10.1093/mnras/stv180},
  \href {https://ui.adsabs.harvard.edu/abs/2015MNRAS.448.3297F} {448, 3297}

\bibitem[\protect\citeauthoryear{{Federrath} \& {Klessen}}{{Federrath} \&
  {Klessen}}{2012}]{2012ApJ...761..156F}
{Federrath} C.,  {Klessen} R.~S.,  2012, \mn@doi [\apj]
  {10.1088/0004-637X/761/2/156}, \href
  {https://ui.adsabs.harvard.edu/abs/2012ApJ...761..156F} {761, 156}

\bibitem[\protect\citeauthoryear{{Federrath} \& {Klessen}}{{Federrath} \&
  {Klessen}}{2013}]{2013ApJ...763...51F}
{Federrath} C.,  {Klessen} R.~S.,  2013, \mn@doi [\apj]
  {10.1088/0004-637X/763/1/51}, \href
  {https://ui.adsabs.harvard.edu/abs/2013ApJ...763...51F} {763, 51}

\bibitem[\protect\citeauthoryear{{Federrath}, {Klessen}  \&
  {Schmidt}}{{Federrath} et~al.}{2008}]{2008ApJ...688L..79F}
{Federrath} C.,  {Klessen} R.~S.,   {Schmidt} W.,  2008, \mn@doi [\apjl]
  {10.1086/595280}, \href
  {https://ui.adsabs.harvard.edu/abs/2008ApJ...688L..79F} {688, L79}

\bibitem[\protect\citeauthoryear{{Federrath}, {Roman-Duval}, {Klessen},
  {Schmidt}  \& {Mac Low}}{{Federrath} et~al.}{2010}]{2010A&A...512A..81F}
{Federrath} C.,  {Roman-Duval} J.,  {Klessen} R.~S.,  {Schmidt} W.,   {Mac Low}
  M.~M.,  2010, \mn@doi [\aap] {10.1051/0004-6361/200912437}, \href
  {https://ui.adsabs.harvard.edu/abs/2010A&A...512A..81F} {512, A81}

\bibitem[\protect\citeauthoryear{{Federrath}, {Chabrier}, {Schober},
  {Banerjee}, {Klessen}  \& {Schleicher}}{{Federrath}
  et~al.}{2011}]{2011PhRvL.107k4504F}
{Federrath} C.,  {Chabrier} G.,  {Schober} J.,  {Banerjee} R.,  {Klessen}
  R.~S.,   {Schleicher} D.~R.~G.,  2011, \mn@doi [\prl]
  {10.1103/PhysRevLett.107.114504}, \href
  {https://ui.adsabs.harvard.edu/abs/2011PhRvL.107k4504F} {107, 114504}

\bibitem[\protect\citeauthoryear{{Federrath} et~al.,}{{Federrath}
  et~al.}{2016}]{2016ApJ...832..143F}
{Federrath} C.,  et~al., 2016, \mn@doi [\apj] {10.3847/0004-637X/832/2/143},
  \href {https://ui.adsabs.harvard.edu/abs/2016ApJ...832..143F} {832, 143}

\bibitem[\protect\citeauthoryear{{Federrath} et~al.,}{{Federrath}
  et~al.}{2017}]{2017IAUS..322..123F}
{Federrath} C.,  et~al., 2017, in {Crocker} R.~M.,  {Longmore} S.~N.,
  {Bicknell} G.~V.,  eds, ~ Vol. 322, The Multi-Messenger Astrophysics of the
  Galactic Centre. pp 123--128 (\mn@eprint {arXiv} {1609.08726}),
  \mn@doi{10.1017/S1743921316012357}

\bibitem[\protect\citeauthoryear{{Federrath}, {Klessen}, {Iapichino}  \&
  {Beattie}}{{Federrath} et~al.}{2021}]{2021NatAs...5..365F}
{Federrath} C.,  {Klessen} R.~S.,  {Iapichino} L.,   {Beattie} J.~R.,  2021,
  \mn@doi [Nature Astronomy] {10.1038/s41550-020-01282-z}, \href
  {https://ui.adsabs.harvard.edu/abs/2021NatAs...5..365F} {5, 365}

\bibitem[\protect\citeauthoryear{{Finn} et~al.,}{{Finn}
  et~al.}{2021}]{2021arXiv210611973F}
{Finn} M.~K.,  et~al., 2021, \mn@doi [\apj] {10.3847/1538-4357/ac090c}, \href
  {https://ui.adsabs.harvard.edu/abs/2021ApJ...917..106F} {917, 106}

\bibitem[\protect\citeauthoryear{{Fischera}}{{Fischera}}{2014}]{2014A&A...565A..24F}
{Fischera} J.,  2014, \mn@doi [\aap] {10.1051/0004-6361/201321417}, \href
  {https://ui.adsabs.harvard.edu/abs/2014A&A...565A..24F} {565, A24}

\bibitem[\protect\citeauthoryear{{Fujimoto} \& {Noguchi}}{{Fujimoto} \&
  {Noguchi}}{1990}]{1990PASJ...42..505F}
{Fujimoto} M.,  {Noguchi} M.,  1990, \pasj, \href
  {https://ui.adsabs.harvard.edu/abs/1990PASJ...42..505F} {42, 505}

\bibitem[\protect\citeauthoryear{{Fukui} et~al.,}{{Fukui}
  et~al.}{2008}]{2008ApJS..178...56F}
{Fukui} Y.,  et~al., 2008, \mn@doi [\apjs] {10.1086/589833}, \href
  {https://ui.adsabs.harvard.edu/abs/2008ApJS..178...56F} {178, 56}

\bibitem[\protect\citeauthoryear{{Fukui} et~al.,}{{Fukui}
  et~al.}{2009}]{2009ApJ...705..144F}
{Fukui} Y.,  et~al., 2009, \mn@doi [\apj] {10.1088/0004-637X/705/1/144}, \href
  {https://ui.adsabs.harvard.edu/abs/2009ApJ...705..144F} {705, 144}

\bibitem[\protect\citeauthoryear{{Fukui} et~al.,}{{Fukui}
  et~al.}{2015}]{2015ApJ...807L...4F}
{Fukui} Y.,  et~al., 2015, \mn@doi [\apjl] {10.1088/2041-8205/807/1/L4}, \href
  {https://ui.adsabs.harvard.edu/abs/2015ApJ...807L...4F} {807, L4}

\bibitem[\protect\citeauthoryear{{Fukui}, {Tsuge}, {Sano}, {Bekki}, {Yozin},
  {Tachihara}  \& {Inoue}}{{Fukui} et~al.}{2017}]{2017PASJ...69L...5F}
{Fukui} Y.,  {Tsuge} K.,  {Sano} H.,  {Bekki} K.,  {Yozin} C.,  {Tachihara} K.,
    {Inoue} T.,  2017, \mn@doi [\pasj] {10.1093/pasj/psx032}, \href
  {https://ui.adsabs.harvard.edu/abs/2017PASJ...69L...5F} {69, L5}

\bibitem[\protect\citeauthoryear{{Fukui} et~al.,}{{Fukui}
  et~al.}{2019}]{2019ApJ...886...14F}
{Fukui} Y.,  et~al., 2019, \mn@doi [\apj] {10.3847/1538-4357/ab4900}, \href
  {https://ui.adsabs.harvard.edu/abs/2019ApJ...886...14F} {886, 14 (F19)}

\bibitem[\protect\citeauthoryear{{Fukui}, {Habe}, {Inoue}, {Enokiya}  \&
  {Tachihara}}{{Fukui} et~al.}{2021}]{2021PASJ...73S...1F}
{Fukui} Y.,  {Habe} A.,  {Inoue} T.,  {Enokiya} R.,   {Tachihara} K.,  2021,
  \mn@doi [\pasj] {10.1093/pasj/psaa103}, \href
  {https://ui.adsabs.harvard.edu/abs/2021PASJ...73S...1F} {73, S1}

\bibitem[\protect\citeauthoryear{{Gaensler}, {Haverkorn}, {Staveley-Smith},
  {Dickey}, {McClure-Griffiths}, {Dickel}  \& {Wolleben}}{{Gaensler}
  et~al.}{2005}]{2005Sci...307.1610G}
{Gaensler} B.~M.,  {Haverkorn} M.,  {Staveley-Smith} L.,  {Dickey} J.~M.,
  {McClure-Griffiths} N.~M.,  {Dickel} J.~R.,   {Wolleben} M.,  2005, \mn@doi
  [Science] {10.1126/science.1108832}, \href
  {https://ui.adsabs.harvard.edu/abs/2005Sci...307.1610G} {307, 1610}

\bibitem[\protect\citeauthoryear{{Galametz} et~al.,}{{Galametz}
  et~al.}{2020}]{2020A&A...643A..63G}
{Galametz} M.,  et~al., 2020, \mn@doi [\aap] {10.1051/0004-6361/202038641},
  \href {https://ui.adsabs.harvard.edu/abs/2020A&A...643A..63G} {643, A63}

\bibitem[\protect\citeauthoryear{{Gatley}, {Becklin}, {Hyland}  \&
  {Jones}}{{Gatley} et~al.}{1981}]{1981MNRAS.197P..17G}
{Gatley} I.,  {Becklin} E.~E.,  {Hyland} A.~R.,   {Jones} T.~J.,  1981, \mn@doi
  [\mnras] {10.1093/mnras/197.1.17P}, \href
  {https://ui.adsabs.harvard.edu/abs/1981MNRAS.197P..17G} {197, 17}

\bibitem[\protect\citeauthoryear{{Ginsburg}, {Federrath}  \&
  {Darling}}{{Ginsburg} et~al.}{2013}]{2013ApJ...779...50G}
{Ginsburg} A.,  {Federrath} C.,   {Darling} J.,  2013, \mn@doi [\apj]
  {10.1088/0004-637X/779/1/50}, \href
  {https://ui.adsabs.harvard.edu/abs/2013ApJ...779...50G} {779, 50}

\bibitem[\protect\citeauthoryear{{Girichidis}, {Konstandin}, {Whitworth}  \&
  {Klessen}}{{Girichidis} et~al.}{2014}]{2014ApJ...781...91G}
{Girichidis} P.,  {Konstandin} L.,  {Whitworth} A.~P.,   {Klessen} R.~S.,
  2014, \mn@doi [\apj] {10.1088/0004-637X/781/2/91}, \href
  {https://ui.adsabs.harvard.edu/abs/2014ApJ...781...91G} {781, 91}

\bibitem[\protect\citeauthoryear{{Girichidis} et~al.,}{{Girichidis}
  et~al.}{2020}]{2020SSRv..216...68G}
{Girichidis} P.,  et~al., 2020, \mn@doi [\ssr] {10.1007/s11214-020-00693-8},
  \href {https://ui.adsabs.harvard.edu/abs/2020SSRv..216...68G} {216, 68}

\bibitem[\protect\citeauthoryear{{Glover} \& {Clark}}{{Glover} \&
  {Clark}}{2012}]{2012MNRAS.426..377G}
{Glover} S. C.~O.,  {Clark} P.~C.,  2012, \mn@doi [\mnras]
  {10.1111/j.1365-2966.2012.21737.x}, \href
  {https://ui.adsabs.harvard.edu/abs/2012MNRAS.426..377G} {426, 377}

\bibitem[\protect\citeauthoryear{{Goldsmith} \& {Langer}}{{Goldsmith} \&
  {Langer}}{1999}]{1999ApJ...517..209G}
{Goldsmith} P.~F.,  {Langer} W.~D.,  1999, \mn@doi [\apj] {10.1086/307195},
  \href {https://ui.adsabs.harvard.edu/abs/1999ApJ...517..209G} {517, 209}

\bibitem[\protect\citeauthoryear{{Goodman}, {Pineda}  \& {Schnee}}{{Goodman}
  et~al.}{2009}]{2009ApJ...692...91G}
{Goodman} A.~A.,  {Pineda} J.~E.,   {Schnee} S.~L.,  2009, \mn@doi [\apj]
  {10.1088/0004-637X/692/1/91}, \href
  {https://ui.adsabs.harvard.edu/abs/2009ApJ...692...91G} {692, 91}

\bibitem[\protect\citeauthoryear{Guicking, Glassmeier, Auster, Delva,
  Motschmann, Narita  \& Zhang}{Guicking et~al.}{2010}]{angeo-28-951-2010}
Guicking L.,  Glassmeier K.-H.,  Auster H.-U.,  Delva M.,  Motschmann U.,
  Narita Y.,   Zhang T.~L.,  2010, \mn@doi [Annales Geophysicae]
  {10.5194/angeo-28-951-2010}, 28, 951

\bibitem[\protect\citeauthoryear{{Hacar}, {Alves}, {Burkert}  \&
  {Goldsmith}}{{Hacar} et~al.}{2016}]{2016A&A...591A.104H}
{Hacar} A.,  {Alves} J.,  {Burkert} A.,   {Goldsmith} P.,  2016, \mn@doi [\aap]
  {10.1051/0004-6361/201527319}, \href
  {https://ui.adsabs.harvard.edu/abs/2016A&A...591A.104H} {591, A104}

\bibitem[\protect\citeauthoryear{{Hansen}, {McKee}  \& {Klein}}{{Hansen}
  et~al.}{2011}]{2011ApJ...738...88H}
{Hansen} C.~E.,  {McKee} C.~F.,   {Klein} R.~I.,  2011, \mn@doi [\apj]
  {10.1088/0004-637X/738/1/88}, \href
  {https://ui.adsabs.harvard.edu/abs/2011ApJ...738...88H} {738, 88}

\bibitem[\protect\citeauthoryear{{Hansen}, {Klein}, {McKee}  \&
  {Fisher}}{{Hansen} et~al.}{2012}]{2012ApJ...747...22H}
{Hansen} C.~E.,  {Klein} R.~I.,  {McKee} C.~F.,   {Fisher} R.~T.,  2012,
  \mn@doi [\apj] {10.1088/0004-637X/747/1/22}, \href
  {https://ui.adsabs.harvard.edu/abs/2012ApJ...747...22H} {747, 22}

\bibitem[\protect\citeauthoryear{{Harris} et~al.,}{{Harris}
  et~al.}{2020}]{2020arXiv200610256H}
{Harris} C.~R.,  et~al., 2020, \mn@doi [\nat] {10.1038/s41586-020-2649-2},
  \href {https://ui.adsabs.harvard.edu/abs/2020arXiv200610256H} {585, 357}

\bibitem[\protect\citeauthoryear{{Heikkila}, {Johansson}  \&
  {Olofsson}}{{Heikkila} et~al.}{1998}]{1998A&A...332..493H}
{Heikkila} A.,  {Johansson} L.~E.~B.,   {Olofsson} H.,  1998, \aap, \href
  {https://ui.adsabs.harvard.edu/abs/1998A&A...332..493H} {332, 493}

\bibitem[\protect\citeauthoryear{{Heydari-Malayeri}, {Rosa}, {Charmandaris},
  {Deharveng}  \& {Zinnecker}}{{Heydari-Malayeri}
  et~al.}{1999}]{1999A&A...352..665H}
{Heydari-Malayeri} M.,  {Rosa} M.~R.,  {Charmandaris} V.,  {Deharveng} L.,
  {Zinnecker} H.,  1999, \aap, \href
  {https://ui.adsabs.harvard.edu/abs/1999A&A...352..665H} {352, 665}

\bibitem[\protect\citeauthoryear{{Heyer} \& {Brunt}}{{Heyer} \&
  {Brunt}}{2012}]{2012MNRAS.420.1562H}
{Heyer} M.~H.,  {Brunt} C.~M.,  2012, \mn@doi [\mnras]
  {10.1111/j.1365-2966.2011.20142.x}, \href
  {https://ui.adsabs.harvard.edu/abs/2012MNRAS.420.1562H} {420, 1562}

\bibitem[\protect\citeauthoryear{{Hopkins}}{{Hopkins}}{2013}]{Hopkins_2013}
{Hopkins} P.~F.,  2013, \mn@doi [\mnras] {10.1093/mnras/stt010}, \href
  {https://ui.adsabs.harvard.edu/abs/2013MNRAS.430.1880H} {430, 1880}

\bibitem[\protect\citeauthoryear{{Hughes}, {Hayashi}  \& {Koyama}}{{Hughes}
  et~al.}{1998}]{1998ApJ...505..732H}
{Hughes} J.~P.,  {Hayashi} I.,   {Koyama} K.,  1998, \mn@doi [\apj]
  {10.1086/306202}, \href
  {https://ui.adsabs.harvard.edu/abs/1998ApJ...505..732H} {505, 732}

\bibitem[\protect\citeauthoryear{{Hughes} et~al.,}{{Hughes}
  et~al.}{2010}]{2010MNRAS.406.2065H}
{Hughes} A.,  et~al., 2010, \mn@doi [\mnras]
  {10.1111/j.1365-2966.2010.16829.x}, \href
  {https://ui.adsabs.harvard.edu/abs/2010MNRAS.406.2065H} {406, 2065}

\bibitem[\protect\citeauthoryear{Hunter}{Hunter}{2007}]{Hunter:2007}
Hunter J.~D.,  2007, \mn@doi [Computing in Science \& Engineering]
  {10.1109/MCSE.2007.55}, 9, 90

\bibitem[\protect\citeauthoryear{{Imara}, {Forbes}  \& {Weaver}}{{Imara}
  et~al.}{2021}]{Imara2021}
{Imara} N.,  {Forbes} J.~C.,   {Weaver} J.~C.,  2021, \mn@doi [\apjl]
  {10.3847/2041-8213/ac194e}, \href
  {https://ui.adsabs.harvard.edu/abs/2021ApJ...918L...3I} {918, L3}

\bibitem[\protect\citeauthoryear{{Immer}, {Kauffmann}, {Pillai}, {Ginsburg}  \&
  {Menten}}{{Immer} et~al.}{2016}]{2016A&A...595A..94I}
{Immer} K.,  {Kauffmann} J.,  {Pillai} T.,  {Ginsburg} A.,   {Menten} K.~M.,
  2016, \mn@doi [\aap] {10.1051/0004-6361/201628777}, \href
  {https://ui.adsabs.harvard.edu/abs/2016A&A...595A..94I} {595, A94}

\bibitem[\protect\citeauthoryear{{Indebetouw}, {Johnson}  \&
  {Conti}}{{Indebetouw} et~al.}{2004}]{2004AJ....128.2206I}
{Indebetouw} R.,  {Johnson} K.~E.,   {Conti} P.,  2004, \mn@doi [\aj]
  {10.1086/424614}, \href
  {https://ui.adsabs.harvard.edu/abs/2004AJ....128.2206I} {128, 2206}

\bibitem[\protect\citeauthoryear{{Indebetouw} et~al.,}{{Indebetouw}
  et~al.}{2013}]{2013ApJ...774...73I}
{Indebetouw} R.,  et~al., 2013, \mn@doi [\apj] {10.1088/0004-637X/774/1/73},
  \href {https://ui.adsabs.harvard.edu/abs/2013ApJ...774...73I} {774, 73}

\bibitem[\protect\citeauthoryear{{Inoue} \& {Fukui}}{{Inoue} \&
  {Fukui}}{2013}]{2013ApJ...774L..31I}
{Inoue} T.,  {Fukui} Y.,  2013, \mn@doi [\apjl] {10.1088/2041-8205/774/2/L31},
  \href {https://ui.adsabs.harvard.edu/abs/2013ApJ...774L..31I} {774, L31}

\bibitem[\protect\citeauthoryear{{Inoue} \& {Inutsuka}}{{Inoue} \&
  {Inutsuka}}{2012}]{2012ApJ...759...35I}
{Inoue} T.,  {Inutsuka} S.-i.,  2012, \mn@doi [\apj]
  {10.1088/0004-637X/759/1/35}, \href
  {https://ui.adsabs.harvard.edu/abs/2012ApJ...759...35I} {759, 35}

\bibitem[\protect\citeauthoryear{{Inoue}, {Hennebelle}, {Fukui}, {Matsumoto},
  {Iwasaki}  \& {Inutsuka}}{{Inoue} et~al.}{2018}]{2018PASJ...70S..53I}
{Inoue} T.,  {Hennebelle} P.,  {Fukui} Y.,  {Matsumoto} T.,  {Iwasaki} K.,
  {Inutsuka} S.-i.,  2018, \mn@doi [\pasj] {10.1093/pasj/psx089}, \href
  {https://ui.adsabs.harvard.edu/abs/2018PASJ...70S..53I} {70, S53}

\bibitem[\protect\citeauthoryear{{Jameson} et~al.,}{{Jameson}
  et~al.}{2018}]{2018ApJ...853..111J}
{Jameson} K.~E.,  et~al., 2018, \mn@doi [\apj] {10.3847/1538-4357/aaa4bb},
  \href {https://ui.adsabs.harvard.edu/abs/2018ApJ...853..111J} {853, 111}

\bibitem[\protect\citeauthoryear{{Jaupart} \& {Chabrier}}{{Jaupart} \&
  {Chabrier}}{2020}]{2020ApJ...903L...2J}
{Jaupart} E.,  {Chabrier} G.,  2020, \mn@doi [\apjl]
  {10.3847/2041-8213/abbda8}, \href
  {https://ui.adsabs.harvard.edu/abs/2020ApJ...903L...2J} {903, L2}

\bibitem[\protect\citeauthoryear{{Jin}, {Salim}, {Federrath}, {Tasker}, {Habe}
  \& {Kainulainen}}{{Jin} et~al.}{2017}]{2017MNRAS.469..383J}
{Jin} K.,  {Salim} D.~M.,  {Federrath} C.,  {Tasker} E.~J.,  {Habe} A.,
  {Kainulainen} J.~T.,  2017, \mn@doi [\mnras] {10.1093/mnras/stx737}, \href
  {https://ui.adsabs.harvard.edu/abs/2017MNRAS.469..383J} {469, 383}

\bibitem[\protect\citeauthoryear{{Johansson}, {Olofsson}, {Hjalmarson},
  {Gredel}  \& {Black}}{{Johansson} et~al.}{1994}]{1994A&A...291...89J}
{Johansson} L.~E.~B.,  {Olofsson} H.,  {Hjalmarson} A.,  {Gredel} R.,   {Black}
  J.~H.,  1994, \aap, \href
  {https://ui.adsabs.harvard.edu/abs/1994A&A...291...89J} {291, 89}

\bibitem[\protect\citeauthoryear{{Kadanoff}}{{Kadanoff}}{2000}]{Kadanoff2000}
{Kadanoff} L.~P.,  2000, Statistical Physics.
~ Vol. 1, World Scientific, Singapore, \mn@doi{https://doi.org/10.1142/4016}

\bibitem[\protect\citeauthoryear{{Kainulainen} \& {Federrath}}{{Kainulainen} \&
  {Federrath}}{2017}]{2017A&A...608L...3K}
{Kainulainen} J.,  {Federrath} C.,  2017, \mn@doi [\aap]
  {10.1051/0004-6361/201731028}, \href
  {https://ui.adsabs.harvard.edu/abs/2017A&A...608L...3K} {608, L3}

\bibitem[\protect\citeauthoryear{{Kainulainen} \& {Tan}}{{Kainulainen} \&
  {Tan}}{2013}]{2013A&A...549A..53K}
{Kainulainen} J.,  {Tan} J.~C.,  2013, \mn@doi [\aap]
  {10.1051/0004-6361/201219526}, \href
  {https://ui.adsabs.harvard.edu/abs/2013A&A...549A..53K} {549, A53}

\bibitem[\protect\citeauthoryear{{Kainulainen}, {Federrath}  \&
  {Henning}}{{Kainulainen} et~al.}{2013}]{2013A&A...553L...8K}
{Kainulainen} J.,  {Federrath} C.,   {Henning} T.,  2013, \mn@doi [\aap]
  {10.1051/0004-6361/201321431}, \href
  {https://ui.adsabs.harvard.edu/abs/2013A&A...553L...8K} {553, L8}

\bibitem[\protect\citeauthoryear{{Kainulainen}, {Federrath}  \&
  {Henning}}{{Kainulainen} et~al.}{2014}]{Kainulainen_2014}
{Kainulainen} J.,  {Federrath} C.,   {Henning} T.,  2014, \mn@doi [Science]
  {10.1126/science.1248724}, \href
  {https://ui.adsabs.harvard.edu/abs/2014Sci...344..183K} {344, 183}

\bibitem[\protect\citeauthoryear{{Khullar}, {Federrath}, {Krumholz}  \&
  {Matzner}}{{Khullar} et~al.}{2021}]{2021MNRAS.tmp.1659K}
{Khullar} S.,  {Federrath} C.,  {Krumholz} M.~R.,   {Matzner} C.~D.,  2021,
  \mn@doi [\mnras] {10.1093/mnras/stab1914}, \href
  {https://ui.adsabs.harvard.edu/abs/2021MNRAS.507.4335K} {507, 4335}

\bibitem[\protect\citeauthoryear{{Kim}, {Staveley-Smith}, {Dopita}, {Sault},
  {Freeman}, {Lee}  \& {Chu}}{{Kim} et~al.}{2003}]{2003ApJS..148..473K}
{Kim} S.,  {Staveley-Smith} L.,  {Dopita} M.~A.,  {Sault} R.~J.,  {Freeman}
  K.~C.,  {Lee} Y.,   {Chu} Y.-H.,  2003, \mn@doi [\apjs] {10.1086/376980},
  \href {https://ui.adsabs.harvard.edu/abs/2003ApJS..148..473K} {148, 473}

\bibitem[\protect\citeauthoryear{{Konstandin}, {Girichidis}, {Federrath}  \&
  {Klessen}}{{Konstandin} et~al.}{2012}]{2012ApJ...761..149K}
{Konstandin} L.,  {Girichidis} P.,  {Federrath} C.,   {Klessen} R.~S.,  2012,
  \mn@doi [\apj] {10.1088/0004-637X/761/2/149}, \href
  {https://ui.adsabs.harvard.edu/abs/2012ApJ...761..149K} {761, 149}

\bibitem[\protect\citeauthoryear{{Konstandin}, {Schmidt}, {Girichidis},
  {Peters}, {Shetty}  \& {Klessen}}{{Konstandin}
  et~al.}{2016}]{2016MNRAS.460.4483K}
{Konstandin} L.,  {Schmidt} W.,  {Girichidis} P.,  {Peters} T.,  {Shetty} R.,
  {Klessen} R.~S.,  2016, \mn@doi [\mnras] {10.1093/mnras/stw1313}, \href
  {https://ui.adsabs.harvard.edu/abs/2016MNRAS.460.4483K} {460, 4483}

\bibitem[\protect\citeauthoryear{{Koribalski} et~al.,}{{Koribalski}
  et~al.}{2020}]{2020Ap&SS.365..118K}
{Koribalski} B.~S.,  et~al., 2020, \mn@doi [\apss]
  {10.1007/s10509-020-03831-4}, \href
  {https://ui.adsabs.harvard.edu/abs/2020Ap&SS.365..118K} {365, 118}

\bibitem[\protect\citeauthoryear{{K{\"o}rtgen}}{{K{\"o}rtgen}}{2020}]{2020MNRAS.497.1263K}
{K{\"o}rtgen} B.,  2020, \mn@doi [\mnras] {10.1093/mnras/staa2028}, \href
  {https://ui.adsabs.harvard.edu/abs/2020MNRAS.497.1263K} {497, 1263}

\bibitem[\protect\citeauthoryear{{K{\"o}rtgen}, {Federrath}  \&
  {Banerjee}}{{K{\"o}rtgen} et~al.}{2017}]{2017MNRAS.472.2496K}
{K{\"o}rtgen} B.,  {Federrath} C.,   {Banerjee} R.,  2017, \mn@doi [\mnras]
  {10.1093/mnras/stx2208}, \href
  {https://ui.adsabs.harvard.edu/abs/2017MNRAS.472.2496K} {472, 2496}

\bibitem[\protect\citeauthoryear{{K{\"o}rtgen}, {Federrath}  \&
  {Banerjee}}{{K{\"o}rtgen} et~al.}{2019}]{2019MNRAS.482.5233K}
{K{\"o}rtgen} B.,  {Federrath} C.,   {Banerjee} R.,  2019, \mn@doi [\mnras]
  {10.1093/mnras/sty3071}, \href
  {https://ui.adsabs.harvard.edu/abs/2019MNRAS.482.5233K} {482, 5233}

\bibitem[\protect\citeauthoryear{{Kowal}, {Lazarian}  \& {Beresnyak}}{{Kowal}
  et~al.}{2007}]{2007ApJ...658..423K}
{Kowal} G.,  {Lazarian} A.,   {Beresnyak} A.,  2007, \mn@doi [\apj]
  {10.1086/511515}, \href
  {https://ui.adsabs.harvard.edu/abs/2007ApJ...658..423K} {658, 423}

\bibitem[\protect\citeauthoryear{{Kritsuk}, {Norman}  \& {Wagner}}{{Kritsuk}
  et~al.}{2011}]{2011ApJ...727L..20K}
{Kritsuk} A.~G.,  {Norman} M.~L.,   {Wagner} R.,  2011, \mn@doi [\apjl]
  {10.1088/2041-8205/727/1/L20}, \href
  {https://ui.adsabs.harvard.edu/abs/2011ApJ...727L..20K} {727, L20}

\bibitem[\protect\citeauthoryear{{Krumholz}}{{Krumholz}}{2014}]{2014MNRAS.437.1662K}
{Krumholz} M.~R.,  2014, \mn@doi [\mnras] {10.1093/mnras/stt2000}, \href
  {https://ui.adsabs.harvard.edu/abs/2014MNRAS.437.1662K} {437, 1662}

\bibitem[\protect\citeauthoryear{{Krumholz} \& {McKee}}{{Krumholz} \&
  {McKee}}{2005}]{2005ApJ...630..250K}
{Krumholz} M.~R.,  {McKee} C.~F.,  2005, \mn@doi [\apj] {10.1086/431734}, \href
  {https://ui.adsabs.harvard.edu/abs/2005ApJ...630..250K} {630, 250}

\bibitem[\protect\citeauthoryear{{Lim}, {Cho}  \& {Yoon}}{{Lim}
  et~al.}{2020}]{2020ApJ...893...75L}
{Lim} J.,  {Cho} J.,   {Yoon} H.,  2020, \mn@doi [\apj]
  {10.3847/1538-4357/ab8066}, \href
  {https://ui.adsabs.harvard.edu/abs/2020ApJ...893...75L} {893, 75}

\bibitem[\protect\citeauthoryear{{Livingston}, {McClure-Griffiths}, {Mao},
  {Ma}, {Gaensler}, {Heald}  \& {Seta}}{{Livingston}
  et~al.}{2021}]{2021MNRAS.502.3814L}
{Livingston} J.~D.,  {McClure-Griffiths} N.~M.,  {Mao} S.~A.,  {Ma} Y.~K.,
  {Gaensler} B.~M.,  {Heald} G.,   {Seta} A.,  2021, \mnras, submitted

\bibitem[\protect\citeauthoryear{{Mandal}, {Federrath}  \&
  {K{\"o}rtgen}}{{Mandal} et~al.}{2020}]{2020MNRAS.493.3098M}
{Mandal} A.,  {Federrath} C.,   {K{\"o}rtgen} B.,  2020, \mn@doi [\mnras]
  {10.1093/mnras/staa468}, \href
  {https://ui.adsabs.harvard.edu/abs/2020MNRAS.493.3098M} {493, 3098}

\bibitem[\protect\citeauthoryear{{Mangum} \& {Shirley}}{{Mangum} \&
  {Shirley}}{2015}]{2015PASP..127..266M}
{Mangum} J.~G.,  {Shirley} Y.~L.,  2015, \mn@doi [\pasp] {10.1086/680323},
  \href {https://ui.adsabs.harvard.edu/abs/2015PASP..127..266M} {127, 266}

\bibitem[\protect\citeauthoryear{{Mao} et~al.,}{{Mao}
  et~al.}{2012}]{2012ApJ...759...25M}
{Mao} S.~A.,  et~al., 2012, \mn@doi [\apj] {10.1088/0004-637X/759/1/25}, \href
  {https://ui.adsabs.harvard.edu/abs/2012ApJ...759...25M} {759, 25}

\bibitem[\protect\citeauthoryear{{Marchal} \&
  {Miville-Desch{\^e}nes}}{{Marchal} \&
  {Miville-Desch{\^e}nes}}{2021}]{2020arXiv201203160M}
{Marchal} A.,  {Miville-Desch{\^e}nes} M.-A.,  2021, \mn@doi [\apj]
  {10.3847/1538-4357/abd108}, \href
  {https://ui.adsabs.harvard.edu/abs/2021ApJ...908..186M} {908, 186}

\bibitem[\protect\citeauthoryear{{Menon}, {Federrath}  \& {Kuiper}}{{Menon}
  et~al.}{2020}]{2020MNRAS.493.4643M}
{Menon} S.~H.,  {Federrath} C.,   {Kuiper} R.,  2020, \mn@doi [\mnras]
  {10.1093/mnras/staa580}, \href
  {https://ui.adsabs.harvard.edu/abs/2020MNRAS.493.4643M} {493, 4643}

\bibitem[\protect\citeauthoryear{{Menon}, {Federrath}, {Klaassen}, {Kuiper}  \&
  {Reiter}}{{Menon} et~al.}{2021}]{2021MNRAS.500.1721M}
{Menon} S.~H.,  {Federrath} C.,  {Klaassen} P.,  {Kuiper} R.,   {Reiter} M.,
  2021, \mn@doi [\mnras] {10.1093/mnras/staa3271}, \href
  {https://ui.adsabs.harvard.edu/abs/2021MNRAS.500.1721M} {500, 1721 (M21)}

\bibitem[\protect\citeauthoryear{{Meynadier}, {Heydari-Malayeri}, {Deharveng},
  {Charmandaris}, {Le Bertre}, {Rosa}, {Schaerer}  \& {Zinnecker}}{{Meynadier}
  et~al.}{2004}]{2004A&A...422..129M}
{Meynadier} F.,  {Heydari-Malayeri} M.,  {Deharveng} L.,  {Charmandaris} V.,
  {Le Bertre} T.,  {Rosa} M.~R.,  {Schaerer} D.,   {Zinnecker} H.,  2004,
  \mn@doi [\aap] {10.1051/0004-6361:20035875}, \href
  {https://ui.adsabs.harvard.edu/abs/2004A&A...422..129M} {422, 129}

\bibitem[\protect\citeauthoryear{{Miesch} \& {Toomre}}{{Miesch} \&
  {Toomre}}{2009}]{2009AnRFM..41..317M}
{Miesch} M.~S.,  {Toomre} J.,  2009, \mn@doi [Annual Review of Fluid Mechanics]
  {10.1146/annurev.fluid.010908.165215}, \href
  {https://ui.adsabs.harvard.edu/abs/2009AnRFM..41..317M} {41, 317}

\bibitem[\protect\citeauthoryear{{Minamidani} et~al.,}{{Minamidani}
  et~al.}{2008}]{2008ApJS..175..485M}
{Minamidani} T.,  et~al., 2008, \mn@doi [\apjs] {10.1086/524038}, \href
  {https://ui.adsabs.harvard.edu/abs/2008ApJS..175..485M} {175, 485}

\bibitem[\protect\citeauthoryear{{Mizuno} et~al.,}{{Mizuno}
  et~al.}{2010}]{2010PASJ...62...51M}
{Mizuno} Y.,  et~al., 2010, \mn@doi [\pasj] {10.1093/pasj/62.1.51}, \href
  {https://ui.adsabs.harvard.edu/abs/2010PASJ...62...51M} {62, 51}

\bibitem[\protect\citeauthoryear{{Mocz} \& {Burkhart}}{{Mocz} \&
  {Burkhart}}{2019}]{Mocz2019}
{Mocz} P.,  {Burkhart} B.,  2019, \mn@doi [\apjl] {10.3847/2041-8213/ab48f6},
  \href {https://ui.adsabs.harvard.edu/abs/2019ApJ...884L..35M} {884, L35}

\bibitem[\protect\citeauthoryear{{Mohapatra} \& {Sharma}}{{Mohapatra} \&
  {Sharma}}{2019}]{2019MNRAS.484.4881M}
{Mohapatra} R.,  {Sharma} P.,  2019, \mn@doi [\mnras] {10.1093/mnras/stz328},
  \href {https://ui.adsabs.harvard.edu/abs/2019MNRAS.484.4881M} {484, 4881}

\bibitem[\protect\citeauthoryear{{Mohapatra}, {Federrath}  \&
  {Sharma}}{{Mohapatra} et~al.}{2020}]{2020MNRAS.493.5838M}
{Mohapatra} R.,  {Federrath} C.,   {Sharma} P.,  2020, \mn@doi [\mnras]
  {10.1093/mnras/staa711}, \href
  {https://ui.adsabs.harvard.edu/abs/2020MNRAS.493.5838M} {493, 5838}

\bibitem[\protect\citeauthoryear{{Molina}, {Glover}, {Federrath}  \&
  {Klessen}}{{Molina} et~al.}{2012}]{2012MNRAS.423.2680M}
{Molina} F.~Z.,  {Glover} S.~C.~O.,  {Federrath} C.,   {Klessen} R.~S.,  2012,
  \mn@doi [\mnras] {10.1111/j.1365-2966.2012.21075.x}, \href
  {https://ui.adsabs.harvard.edu/abs/2012MNRAS.423.2680M} {423, 2680}

\bibitem[\protect\citeauthoryear{{Nakajima} et~al.,}{{Nakajima}
  et~al.}{2005}]{2005AJ....129..776N}
{Nakajima} Y.,  et~al., 2005, \mn@doi [\aj] {10.1086/426917}, \href
  {https://ui.adsabs.harvard.edu/abs/2005AJ....129..776N} {129, 776}

\bibitem[\protect\citeauthoryear{{Nayak}, {Meixner}, {Fukui}, {Tachihara},
  {Onishi}, {Saigo}, {Tokuda}  \& {Harada}}{{Nayak}
  et~al.}{2018}]{2018ApJ...854..154N}
{Nayak} O.,  {Meixner} M.,  {Fukui} Y.,  {Tachihara} K.,  {Onishi} T.,  {Saigo}
  K.,  {Tokuda} K.,   {Harada} R.,  2018, \mn@doi [\apj]
  {10.3847/1538-4357/aaab5f}, \href
  {https://ui.adsabs.harvard.edu/abs/2018ApJ...854..154N} {854, 154}

\bibitem[\protect\citeauthoryear{{Nolan}, {Federrath}  \& {Sutherland}}{{Nolan}
  et~al.}{2015}]{2015MNRAS.451.1380N}
{Nolan} C.~A.,  {Federrath} C.,   {Sutherland} R.~S.,  2015, \mn@doi [\mnras]
  {10.1093/mnras/stv1030}, \href
  {https://ui.adsabs.harvard.edu/abs/2015MNRAS.451.1380N} {451, 1380}

\bibitem[\protect\citeauthoryear{{Offner} \& {Arce}}{{Offner} \&
  {Arce}}{2014}]{2014ApJ...784...61O}
{Offner} S. S.~R.,  {Arce} H.~G.,  2014, \mn@doi [\apj]
  {10.1088/0004-637X/784/1/61}, \href
  {https://ui.adsabs.harvard.edu/abs/2014ApJ...784...61O} {784, 61}

\bibitem[\protect\citeauthoryear{{Offner} \& {Chaban}}{{Offner} \&
  {Chaban}}{2017}]{2017ApJ...847..104O}
{Offner} S. S.~R.,  {Chaban} J.,  2017, \mn@doi [\apj]
  {10.3847/1538-4357/aa8996}, \href
  {https://ui.adsabs.harvard.edu/abs/2017ApJ...847..104O} {847, 104}

\bibitem[\protect\citeauthoryear{{Offner} \& {Liu}}{{Offner} \&
  {Liu}}{2018}]{2018NatAs...2..896O}
{Offner} S. S.~R.,  {Liu} Y.,  2018, \mn@doi [Nature Astronomy]
  {10.1038/s41550-018-0566-1}, \href
  {https://ui.adsabs.harvard.edu/abs/2018NatAs...2..896O} {2, 896}

\bibitem[\protect\citeauthoryear{{Oliphant}}{{Oliphant}}{2006}]{oliphant2006guide}
{Oliphant} T.~E.,  2006, A guide to NumPy.
~ Vol. 1, Trelgol Publishing USA

\bibitem[\protect\citeauthoryear{{Orkisz} et~al.,}{{Orkisz}
  et~al.}{2017}]{2017A&A...599A..99O}
{Orkisz} J.~H.,  et~al., 2017, \mn@doi [\aap] {10.1051/0004-6361/201629220},
  \href {https://ui.adsabs.harvard.edu/abs/2017A&A...599A..99O} {599, A99}

\bibitem[\protect\citeauthoryear{{Ossenkopf} \& {Mac Low}}{{Ossenkopf} \& {Mac
  Low}}{2002}]{2002A&A...390..307O}
{Ossenkopf} V.,  {Mac Low} M.~M.,  2002, \mn@doi [\aap]
  {10.1051/0004-6361:20020629}, \href
  {https://ui.adsabs.harvard.edu/abs/2002A&A...390..307O} {390, 307}

\bibitem[\protect\citeauthoryear{{Ossenkopf-Okada}, {Csengeri}, {Schneider},
  {Federrath}  \& {Klessen}}{{Ossenkopf-Okada}
  et~al.}{2016}]{2016A&A...590A.104O}
{Ossenkopf-Okada} V.,  {Csengeri} T.,  {Schneider} N.,  {Federrath} C.,
  {Klessen} R.~S.,  2016, \mn@doi [\aap] {10.1051/0004-6361/201628095}, \href
  {https://ui.adsabs.harvard.edu/abs/2016A&A...590A.104O} {590, A104}

\bibitem[\protect\citeauthoryear{{Ossenkopf}, {Krips}  \&
  {Stutzki}}{{Ossenkopf} et~al.}{2008}]{Ossenkopf_2008}
{Ossenkopf} V.,  {Krips} M.,   {Stutzki} J.,  2008, \mn@doi [\aap]
  {10.1051/0004-6361:20079106}, \href
  {https://ui.adsabs.harvard.edu/abs/2008A&A...485..917O} {485, 917}

\bibitem[\protect\citeauthoryear{{Padoan} \& {Nordlund}}{{Padoan} \&
  {Nordlund}}{2011}]{2011ApJ...730...40P}
{Padoan} P.,  {Nordlund} {\r{A}}.,  2011, \mn@doi [\apj]
  {10.1088/0004-637X/730/1/40}, \href
  {https://ui.adsabs.harvard.edu/abs/2011ApJ...730...40P} {730, 40}

\bibitem[\protect\citeauthoryear{{Padoan}, {Nordlund}  \& {Jones}}{{Padoan}
  et~al.}{1997a}]{1997MNRAS.288..145P}
{Padoan} P.,  {Nordlund} A.,   {Jones} B. J.~T.,  1997a, \mn@doi [\mnras]
  {10.1093/mnras/288.1.145}, \href
  {https://ui.adsabs.harvard.edu/abs/1997MNRAS.288..145P} {288, 145}

\bibitem[\protect\citeauthoryear{{Padoan}, {Jones}  \& {Nordlund}}{{Padoan}
  et~al.}{1997b}]{1997ApJ...474..730P}
{Padoan} P.,  {Jones} B. J.~T.,   {Nordlund} {\r{A}}.~P.,  1997b, \mn@doi
  [\apj] {10.1086/303482}, \href
  {https://ui.adsabs.harvard.edu/abs/1997ApJ...474..730P} {474, 730}

\bibitem[\protect\citeauthoryear{{Pan} \& {Padoan}}{{Pan} \&
  {Padoan}}{2009}]{2009ApJ...692..594P}
{Pan} L.,  {Padoan} P.,  2009, \mn@doi [\apj] {10.1088/0004-637X/692/1/594},
  \href {https://ui.adsabs.harvard.edu/abs/2009ApJ...692..594P} {692, 594}

\bibitem[\protect\citeauthoryear{{Pan}, {Padoan}, {Haugb{\o}lle}  \&
  {Nordlund}}{{Pan} et~al.}{2016}]{2016ApJ...825...30P}
{Pan} L.,  {Padoan} P.,  {Haugb{\o}lle} T.,   {Nordlund} {\r{A}}.,  2016,
  \mn@doi [\apj] {10.3847/0004-637X/825/1/30}, \href
  {https://ui.adsabs.harvard.edu/abs/2016ApJ...825...30P} {825, 30}

\bibitem[\protect\citeauthoryear{{Passot} \& {V{\'a}zquez-Semadeni}}{{Passot}
  \& {V{\'a}zquez-Semadeni}}{1998}]{1998PhRvE..58.4501P}
{Passot} T.,  {V{\'a}zquez-Semadeni} E.,  1998, \mn@doi [\pre]
  {10.1103/PhysRevE.58.4501}, \href
  {https://ui.adsabs.harvard.edu/abs/1998PhRvE..58.4501P} {58, 4501}

\bibitem[\protect\citeauthoryear{{Pietrzy{\'n}ski} et~al.,}{{Pietrzy{\'n}ski}
  et~al.}{2019}]{2019Natur.567..200P}
{Pietrzy{\'n}ski} G.,  et~al., 2019, \mn@doi [\nat]
  {10.1038/s41586-019-0999-4}, \href
  {https://ui.adsabs.harvard.edu/abs/2019Natur.567..200P} {567, 200}

\bibitem[\protect\citeauthoryear{{Pingel}, {Dempsey}, {McClure-Griffiths}  \&
  {et al.,}}{{Pingel} et~al.}{2021}]{2021pingel}
{Pingel} N.~M.,  {Dempsey} J.,  {McClure-Griffiths} N.~M.,   {et al.,} 2021,
  \pasa, submitted

\bibitem[\protect\citeauthoryear{{Robertson} \& {Goldreich}}{{Robertson} \&
  {Goldreich}}{2018}]{Robertson2018}
{Robertson} B.,  {Goldreich} P.,  2018, \mn@doi [\apj]
  {10.3847/1538-4357/aaa89e}, \href
  {https://ui.adsabs.harvard.edu/abs/2018ApJ...854...88R} {854, 88}

\bibitem[\protect\citeauthoryear{{Roman-Duval} et~al.,}{{Roman-Duval}
  et~al.}{2014}]{2014ApJ...797...86R}
{Roman-Duval} J.,  et~al., 2014, \mn@doi [\apj] {10.1088/0004-637X/797/2/86},
  \href {https://ui.adsabs.harvard.edu/abs/2014ApJ...797...86R} {797, 86}

\bibitem[\protect\citeauthoryear{{Rosen}, {Offner}, {Sadavoy}, {Bhandare},
  {V{\'a}zquez-Semadeni}  \& {Ginsburg}}{{Rosen}
  et~al.}{2020}]{2020SSRv..216...62R}
{Rosen} A.~L.,  {Offner} S. S.~R.,  {Sadavoy} S.~I.,  {Bhandare} A.,
  {V{\'a}zquez-Semadeni} E.,   {Ginsburg} A.,  2020, \mn@doi [\ssr]
  {10.1007/s11214-020-00688-5}, \href
  {https://ui.adsabs.harvard.edu/abs/2020SSRv..216...62R} {216, 62}

\bibitem[\protect\citeauthoryear{{Saigo} et~al.,}{{Saigo}
  et~al.}{2017}]{2017ApJ...835..108S}
{Saigo} K.,  et~al., 2017, \mn@doi [\apj] {10.3847/1538-4357/835/1/108}, \href
  {https://ui.adsabs.harvard.edu/abs/2017ApJ...835..108S} {835, 108}

\bibitem[\protect\citeauthoryear{{Salim}, {Federrath}  \& {Kewley}}{{Salim}
  et~al.}{2015}]{2015ApJ...806L..36S}
{Salim} D.~M.,  {Federrath} C.,   {Kewley} L.~J.,  2015, \mn@doi [\apjl]
  {10.1088/2041-8205/806/2/L36}, \href
  {https://ui.adsabs.harvard.edu/abs/2015ApJ...806L..36S} {806, L36}

\bibitem[\protect\citeauthoryear{{Scannapieco} \& {Safarzadeh}}{{Scannapieco}
  \& {Safarzadeh}}{2018}]{2018ApJ...865L..14S}
{Scannapieco} E.,  {Safarzadeh} M.,  2018, \mn@doi [\apjl]
  {10.3847/2041-8213/aae1f9}, \href
  {https://ui.adsabs.harvard.edu/abs/2018ApJ...865L..14S} {865, L14}

\bibitem[\protect\citeauthoryear{{Schenck}, {Park}  \& {Post}}{{Schenck}
  et~al.}{2016}]{2016AJ....151..161S}
{Schenck} A.,  {Park} S.,   {Post} S.,  2016, \mn@doi [\aj]
  {10.3847/0004-6256/151/6/161}, \href
  {https://ui.adsabs.harvard.edu/abs/2016AJ....151..161S} {151, 161}

\bibitem[\protect\citeauthoryear{{Schneider} et~al.,}{{Schneider}
  et~al.}{2011}]{2011A&A...529A...1S}
{Schneider} N.,  et~al., 2011, \mn@doi [\aap] {10.1051/0004-6361/200913884},
  \href {https://ui.adsabs.harvard.edu/abs/2011A&A...529A...1S} {529, A1}

\bibitem[\protect\citeauthoryear{{Schneider} et~al.,}{{Schneider}
  et~al.}{2013}]{2013ApJ...766L..17S}
{Schneider} N.,  et~al., 2013, \mn@doi [\apjl] {10.1088/2041-8205/766/2/L17},
  \href {https://ui.adsabs.harvard.edu/abs/2013ApJ...766L..17S} {766, L17}

\bibitem[\protect\citeauthoryear{{Schneider} et~al.,}{{Schneider}
  et~al.}{2015a}]{2015MNRAS.453L..41S}
{Schneider} N.,  et~al., 2015a, \mn@doi [\mnras] {10.1093/mnrasl/slv101}, \href
  {https://ui.adsabs.harvard.edu/abs/2015MNRAS.453L..41S} {453, L41}

\bibitem[\protect\citeauthoryear{{Schneider} et~al.,}{{Schneider}
  et~al.}{2015b}]{2015A&A...578A..29S}
{Schneider} N.,  et~al., 2015b, \mn@doi [\aap] {10.1051/0004-6361/201424375},
  \href {https://ui.adsabs.harvard.edu/abs/2015A&A...578A..29S} {578, A29}

\bibitem[\protect\citeauthoryear{{Schneider} et~al.,}{{Schneider}
  et~al.}{2016}]{2016A&A...587A..74S}
{Schneider} N.,  et~al., 2016, \mn@doi [\aap] {10.1051/0004-6361/201527144},
  \href {https://ui.adsabs.harvard.edu/abs/2016A&A...587A..74S} {587, A74}

\bibitem[\protect\citeauthoryear{{Sch{\"o}ier}, {van der Tak}, {van Dishoeck}
  \& {Black}}{{Sch{\"o}ier} et~al.}{2005}]{2005A&A...432..369S}
{Sch{\"o}ier} F.~L.,  {van der Tak} F.~F.~S.,  {van Dishoeck} E.~F.,   {Black}
  J.~H.,  2005, \mn@doi [\aap] {10.1051/0004-6361:20041729}, \href
  {https://ui.adsabs.harvard.edu/abs/2005A&A...432..369S} {432, 369}

\bibitem[\protect\citeauthoryear{{Seale}, {Looney}, {Chu}, {Gruendl}, {Brandl},
  {Chen}, {Brandner}  \& {Blake}}{{Seale} et~al.}{2009}]{2009ApJ...699..150S}
{Seale} J.~P.,  {Looney} L.~W.,  {Chu} Y.-H.,  {Gruendl} R.~A.,  {Brandl} B.,
  {Chen} C. H.~R.,  {Brandner} W.,   {Blake} G.~A.,  2009, \mn@doi [\apj]
  {10.1088/0004-637X/699/1/150}, \href
  {https://ui.adsabs.harvard.edu/abs/2009ApJ...699..150S} {699, 150}

\bibitem[\protect\citeauthoryear{{Sharda}, {Federrath}, {da Cunha}, {Swinbank}
  \& {Dye}}{{Sharda} et~al.}{2018}]{2018MNRAS.477.4380S}
{Sharda} P.,  {Federrath} C.,  {da Cunha} E.,  {Swinbank} A.~M.,   {Dye} S.,
  2018, \mn@doi [\mnras] {10.1093/mnras/sty886}, \href
  {https://ui.adsabs.harvard.edu/abs/2018MNRAS.477.4380S} {477, 4380}

\bibitem[\protect\citeauthoryear{{Sharda} et~al.,}{{Sharda}
  et~al.}{2019a}]{2019MNRAS.487.4305S}
{Sharda} P.,  et~al., 2019a, \mn@doi [\mnras] {10.1093/mnras/stz1543}, \href
  {https://ui.adsabs.harvard.edu/abs/2019MNRAS.487.4305S} {487, 4305}

\bibitem[\protect\citeauthoryear{{Sharda}, {Krumholz}  \& {Federrath}}{{Sharda}
  et~al.}{2019b}]{2019MNRAS.490..513S}
{Sharda} P.,  {Krumholz} M.~R.,   {Federrath} C.,  2019b, \mn@doi [\mnras]
  {10.1093/mnras/stz2618}, \href
  {https://ui.adsabs.harvard.edu/abs/2019MNRAS.490..513S} {490, 513}

\bibitem[\protect\citeauthoryear{{Sharda}, {Federrath}  \& {Krumholz}}{{Sharda}
  et~al.}{2020}]{2020MNRAS.497..336S}
{Sharda} P.,  {Federrath} C.,   {Krumholz} M.~R.,  2020, \mn@doi [\mnras]
  {10.1093/mnras/staa1926}, \href
  {https://ui.adsabs.harvard.edu/abs/2020MNRAS.497..336S} {497, 336}

\bibitem[\protect\citeauthoryear{{Sharda}, {Federrath}, {Krumholz}  \&
  {Schleicher}}{{Sharda} et~al.}{2021}]{2021MNRAS.503.2014S}
{Sharda} P.,  {Federrath} C.,  {Krumholz} M.~R.,   {Schleicher} D. R.~G.,
  2021, \mn@doi [\mnras] {10.1093/mnras/stab531}, \href
  {https://ui.adsabs.harvard.edu/abs/2021MNRAS.503.2014S} {503, 2014}

\bibitem[\protect\citeauthoryear{{Sobolev}}{{Sobolev}}{1960}]{1960mes..book.....S}
{Sobolev} V.~V.,  1960, {Moving envelopes of stars}.
Harvard University Press

\bibitem[\protect\citeauthoryear{{Squire} \& {Hopkins}}{{Squire} \&
  {Hopkins}}{2017}]{2017MNRAS.471.3753S}
{Squire} J.,  {Hopkins} P.~F.,  2017, \mn@doi [\mnras] {10.1093/mnras/stx1817},
  \href {https://ui.adsabs.harvard.edu/abs/2017MNRAS.471.3753S} {471, 3753}

\bibitem[\protect\citeauthoryear{{Stewart} \& {Federrath}}{{Stewart} \&
  {Federrath}}{2021}]{2021stewart}
{Stewart} M.,  {Federrath} C.,  2021, \mnras, submitted

\bibitem[\protect\citeauthoryear{{Struck} \& {Smith}}{{Struck} \&
  {Smith}}{1999}]{1999ApJ...527..673S}
{Struck} C.,  {Smith} D.~C.,  1999, \mn@doi [\apj] {10.1086/308112}, \href
  {https://ui.adsabs.harvard.edu/abs/1999ApJ...527..673S} {527, 673}

\bibitem[\protect\citeauthoryear{{Tang} et~al.,}{{Tang}
  et~al.}{2017}]{2017A&A...600A..16T}
{Tang} X.~D.,  et~al., 2017, \mn@doi [\aap] {10.1051/0004-6361/201630183},
  \href {https://ui.adsabs.harvard.edu/abs/2017A&A...600A..16T} {600, A16}

\bibitem[\protect\citeauthoryear{{Tang} et~al.,}{{Tang}
  et~al.}{2021}]{2021arXiv210810519T}
{Tang} X.~D.,  et~al., 2021, arXiv e-prints, \href
  {https://ui.adsabs.harvard.edu/abs/2021arXiv210810519T} {p. arXiv:2108.10519}

\bibitem[\protect\citeauthoryear{{Testor}, {Lemaire}, {Field}  \&
  {Diana}}{{Testor} et~al.}{2006}]{2006A&A...453..517T}
{Testor} G.,  {Lemaire} J.~L.,  {Field} D.,   {Diana} S.,  2006, \mn@doi [\aap]
  {10.1051/0004-6361:20054697}, \href
  {https://ui.adsabs.harvard.edu/abs/2006A&A...453..517T} {453, 517}

\bibitem[\protect\citeauthoryear{{Testor}, {Lemaire}, {Kristensen}, {Field}  \&
  {Diana}}{{Testor} et~al.}{2007}]{2007A&A...469..459T}
{Testor} G.,  {Lemaire} J.~L.,  {Kristensen} L.~E.,  {Field} D.,   {Diana} S.,
  2007, \mn@doi [\aap] {10.1051/0004-6361:20066926}, \href
  {https://ui.adsabs.harvard.edu/abs/2007A&A...469..459T} {469, 459}

\bibitem[\protect\citeauthoryear{{Tokuda} et~al.,}{{Tokuda}
  et~al.}{2019}]{2019ApJ...886...15T}
{Tokuda} K.,  et~al., 2019, \mn@doi [\apj] {10.3847/1538-4357/ab48ff}, \href
  {https://ui.adsabs.harvard.edu/abs/2019ApJ...886...15T} {886, 15}

\bibitem[\protect\citeauthoryear{{Tokuda} et~al.,}{{Tokuda}
  et~al.}{2021}]{2021arXiv210809018T}
{Tokuda} K.,  et~al., 2021, arXiv e-prints, \href
  {https://ui.adsabs.harvard.edu/abs/2021arXiv210809018T} {p. arXiv:2108.09018}

\bibitem[\protect\citeauthoryear{{Virtanen} et~al.,}{{Virtanen}
  et~al.}{2020}]{2020NatMe..17..261V}
{Virtanen} P.,  et~al., 2020, \mn@doi [Nature Methods]
  {10.1038/s41592-019-0686-2}, \href
  {https://ui.adsabs.harvard.edu/abs/2020NatMe..17..261V} {17, 261}

\bibitem[\protect\citeauthoryear{{Wang}, {Chin}, {Henkel}, {Whiteoak}  \&
  {Cunningham}}{{Wang} et~al.}{2009}]{2009ApJ...690..580W}
{Wang} M.,  {Chin} Y.~N.,  {Henkel} C.,  {Whiteoak} J.~B.,   {Cunningham} M.,
  2009, \mn@doi [\apj] {10.1088/0004-637X/690/1/580}, \href
  {https://ui.adsabs.harvard.edu/abs/2009ApJ...690..580W} {690, 580}

\bibitem[\protect\citeauthoryear{{Wilson} \& {Rood}}{{Wilson} \&
  {Rood}}{1994}]{1994ARA&A..32..191W}
{Wilson} T.~L.,  {Rood} R.,  1994, \mn@doi [\araa]
  {10.1146/annurev.aa.32.090194.001203}, \href
  {https://ui.adsabs.harvard.edu/abs/1994ARA&A..32..191W} {32, 191}

\bibitem[\protect\citeauthoryear{{Yan} et~al.,}{{Yan}
  et~al.}{2019}]{2019ApJ...877..154Y}
{Yan} Y.~T.,  et~al., 2019, \mn@doi [\apj] {10.3847/1538-4357/ab17d6}, \href
  {https://ui.adsabs.harvard.edu/abs/2019ApJ...877..154Y} {877, 154}

\bibitem[\protect\citeauthoryear{{Yuan}, {Krumholz}  \& {Burkhart}}{{Yuan}
  et~al.}{2020}]{2020MNRAS.498.2440Y}
{Yuan} Y.,  {Krumholz} M.~R.,   {Burkhart} B.,  2020, \mn@doi [\mnras]
  {10.1093/mnras/staa2432}, \href
  {https://ui.adsabs.harvard.edu/abs/2020MNRAS.498.2440Y} {498, 2440}

\bibitem[\protect\citeauthoryear{{Zhuravleva} et~al.,}{{Zhuravleva}
  et~al.}{2014}]{2014Natur.515...85Z}
{Zhuravleva} I.,  et~al., 2014, \mn@doi [\nat] {10.1038/nature13830}, \href
  {https://ui.adsabs.harvard.edu/abs/2014Natur.515...85Z} {515, 85}

\bibitem[\protect\citeauthoryear{{de Jong}, {Chu}  \& {Dalgarno}}{{de Jong}
  et~al.}{1975}]{1975ApJ...199...69D}
{de Jong} T.,  {Chu} S.,   {Dalgarno} A.,  1975, \mn@doi [\apj]
  {10.1086/153665}, \href
  {https://ui.adsabs.harvard.edu/abs/1975ApJ...199...69D} {199, 69}

\bibitem[\protect\citeauthoryear{{de Jong}, {Boland}  \& {Dalgarno}}{{de Jong}
  et~al.}{1980}]{1980A&A....91...68D}
{de Jong} T.,  {Boland} W.,   {Dalgarno} A.,  1980, \aap, \href
  {https://ui.adsabs.harvard.edu/abs/1980A&A....91...68D} {91, 68}

\bibitem[\protect\citeauthoryear{{van der Tak}, {Black}, {Sch{\"o}ier},
  {Jansen}  \& {van Dishoeck}}{{van der Tak}
  et~al.}{2007}]{2007A&A...468..627V}
{van der Tak} F.~F.~S.,  {Black} J.~H.,  {Sch{\"o}ier} F.~L.,  {Jansen} D.~J.,
   {van Dishoeck} E.~F.,  2007, \mn@doi [\aap] {10.1051/0004-6361:20066820},
  \href {https://ui.adsabs.harvard.edu/abs/2007A&A...468..627V} {468, 627}

\bibitem[\protect\citeauthoryear{{van der Tak}, {Lique}, {Faure}, {Black}  \&
  {van Dishoeck}}{{van der Tak} et~al.}{2020}]{2020Atoms...8...15V}
{van der Tak} F. F.~S.,  {Lique} F.,  {Faure} A.,  {Black} J.~H.,   {van
  Dishoeck} E.~F.,  2020, \mn@doi [Atoms] {10.3390/atoms8020015}, \href
  {https://ui.adsabs.harvard.edu/abs/2020Atoms...8...15V} {8, 15}

\makeatother
\end{thebibliography}

%%%%%%%%%%%%%%%%% APPENDICES %%%%%%%%%%%%%%%%%%%%%

\appendix
\section{Mocz and Burkhart (2019) non-lognormal density model}
\label{s:app_moczburkhart}
\subsection{Model description}
\citet{Mocz2019} model the volume-weighted density PDF of hydrodynamical density fluctuations using a Markovian framework. They construct a Langevin model
\begin{align}
s(t + dt) &= s(t) + A(s)\diff{t} + \mathcal{N}(0,1)\sqrt{D(s)\diff{t}},\\
A(s) &= -\frac{s-s_{0}}{\tau_{A}}\left[1 + H(s-s_{0})\frac{3f}{2} \right], \\
D(s) &= \frac{2\sigma_{s}^2}{\tau_{0}},
\end{align}
where $A(s)$ is the deterministic, or advective term in the model, $D(s)$ is the stochastic, or diffusive term in the model and $\mathcal{N}(0,1)$ is a standard normal distribution. The stochastic term, $D(s)$, contains the turbulent fluctuations $\sigma_s^2$, which have dynamical time-scale $\tau_{0} = \ell_{\rm 0}/(c_s\M)$, where $\ell_{\rm 0}$ is the turbulent driving scale. The deterministic term, $A(s)$, encodes how logarithmic density fluctuates about the mean dynamically on timescales $\tau_{A} = \tau_{A,0} / [1 + H(s-s_{0,V})\frac{3f}{2}]$, where $H(s-s_{0})$ is the Heaviside function. This encodes how high-density structures, such as the density contrast caused by a shock, live on shorter times scales than the rest of the density fluctuations in the fluid \citep{Robertson2018}. For $s > s_{0}$ the timescale is reduced by $\tau_{A} = 3f/2$, and hence $f$ becomes the fitting parameter for how much shorter the dynamical timescales are for the over-dense regions. The PDF of \citet{Mocz2019}'s Langevin model defines a steady-state solution to the Fokker-Planck equation \citep{Kadanoff2000}, which has a solution of the form,
\begin{align} \label{eq:MB_pdf}
        p_{\rm MB}(s) &\propto \exp\left\{ -\frac{(s-s_{0})^2[1 + f(s - s_{0})H(s-s_{0})]}{2\sigma_{s}^2 (\tau_{A}/\tau_{0})} \right\}.
\end{align}
The asymmetric dynamical timescale of the low- and high-density objects in the cloud encodes a non-Gaussian 3$^{\rm rd}$ moment (skewness) into the PDF through the $\mathcal{O}([s-s_0]^3)$ term in the exponential. 

\subsection{Fitting and Results}
We fit \autoref{eq:MB_pdf} using the same bootstrapping method we used to fit the \citet{Hopkins_2013} PDF model in \autoref{s:hopkinsPDF}. Each fit in the bootstrapped sample preserves probability, $\int \diff{\eta} \, p_{\rm MB} (\eta) = 1$, $\int \diff{\eta} \, e^{\eta} p_{\rm MB} (\eta) = N_0$. Our fit results, $\sigma_{N/N_0} = 0.56 \pm 0.37$ and $f = 3.15 \pm 0.42$, are consistent within $1\sigma$ variations to the results quoted in \autoref{tab:tab1}, and hence our measurements of the density dispersion are robust to different non-lognormal density PDF models.

% Don't change these lines
\bsp	% typesetting comment
\label{lastpage}
\end{document}